\documentclass[apj,twocolumn,appendixfloats.twocolappendix,numberedappendix,colorlinks,]{openjournal}    
\usepackage{hyperref}
\usepackage{nameref} 
\hypersetup{linkcolor=red,citecolor=Blue,filecolor=cyan,urlcolor=magenta}
\usepackage{tabularx}
\usepackage{longtable}
\expandafter\let\expandafter\oldlongtablestar\csname longtable*\endcsname
\expandafter\let\expandafter\oldendlongtablestar\csname endlongtable*\endcsname
\expandafter\let\csname longtable*\endcsname\relax
\expandafter\let\csname endlongtable*\endcsname\relax
\usepackage{booktabs,caption}
\expandafter\let\csname longtable*\endcsname\oldlongtablestar
\expandafter\let\csname endlongtable*\endcsname\oldendlongtablestar
\usepackage[flushleft]{threeparttable}
\usepackage[dvipsnames]{xcolor}
\usepackage{soul}
\usepackage{amsmath}
\usepackage{textcomp,gensymb}
\usepackage{orcidlink}
\usepackage[T1]{fontenc}
\usepackage{rotating}
\usepackage{sidecap} 
\sidecaptionvpos{figure}{m}
\usepackage{subcaption}
\usepackage{placeins}
\usepackage{wrapfig}

\renewcommand{\baselinestretch}{1.05} 
\newcommand{\avgsnr}{$\langle$SNR$\rangle$}{}

\newcommand{\sblim}{$\mathcal{M}_{\rm{lim}}$}{}
\newcommand{\sblimm}{\mathcal{M}_{\rm{lim}}}

\newcommand{\reseff}{$\mathcal{R}_{\rm{eff}}$}{}

{}
\newcommand{\ts}{\textsuperscript}

\newcommand{\new}{\textcolor{black}}

\defcitealias{statmorph}{R19}
\defcitealias{Abraham2003}{A03}
\defcitealias{Abraham1996}{A96}
\defcitealias{Lotz2004}{L04}
\defcitealias{Lotz2008}{L08}
\defcitealias{Kent1985}{K85}
\defcitealias{Bershady2000}{B00}
\defcitealias{Conselice2003}{C03}
\defcitealias{Conselice2000}{C00}
\defcitealias{Pawlik2016}{Pa16}
\defcitealias{Peth2016}{Pe16}
\defcitealias{Pawlik2018}{P18}
\defcitealias{Sazonova2024}{S24}
\defcitealias{Sazonova2025}{S25}
\defcitealias{Freeman2013}{F13}
\defcitealias{Wen2016}{W16}
\defcitealias{Petrosian1976}{P76}
\defcitealias{galfit}{P02}
\defcitealias{Sersic1963}{S63}
\defcitealias{Graham2005}{G05}

\shorttitle{Morphological metrics and their biases}
\shortauthors{Sazonova et al.}
\graphicspath{{./}{}}

\begin{document}

\title{The Hubble Sequence in JWST CEERS from unbiased galaxy morphologies}

\author{\vspace{-1.3cm}Elizaveta Sazonova\,\orcidlink{0000-0001-6245-5121}$^{1,2}$}
\author{Cameron R. Morgan\,\orcidlink{0009-0009-2522-3685}$^{1,2}$}
\author{Michael Balogh\,\orcidlink{0000-0003-4849-9536}$^{1,2}$}

\affiliation{$^{1}$Waterloo Centre for Astrophysics, University of Waterloo, Waterloo, ON, N2L 3G1 Canada}\email{liza.sazonova@uwaterloo.ca}
\affiliation{$^{2}$Department of Physics and Astronomy, University of Waterloo, Waterloo, ON N2L 3G1, Canada}

\begin{abstract}

Whether the ``Hubble sequence'' of galaxy morphologies exists up to $z\sim4$ is still disputed, and one of the challenges is characterizing galaxy structure consistently across a wide range of redshifts. To enable a fair comparison across cosmic time, we constructed "absolute" images of galaxies spanning $0.15 < z < 4.5$ and $8 < \log M_\star < 11$ from HST CANDELS and JWST CEERS surveys, by matching the effective resolution and surface brightness limit of galaxies, accounting for cosmological dimming and evolution in size and mass-to-light ratio. 

We measured the structural parameters of 2825 galaxies and used the UMAP technique to study the evolution of the morphological phase space. We find a continuous sequence spanning late-type to early-type galaxies, with no redshift gradient -- indicating that a Hubble-like sequence is established by z $\sim$ 4. We show that our approach recovers a cleaner separation between early- and late-type galaxies than visual classifications. By tracing progenitors using empirical mass assembly histories, we find that progenitors of low-mass galaxies are predominantly star-forming disks at all epochs. Progenitors of massive galaxies follow two distinct paths: a stable star-forming disk population with little structural evolution, and an early-type population that builds up rapidly from irregular progenitors and quenches within a few Gyr, consistent with a compaction-driven quenching scenario. 
\end{abstract}


\maketitle

\section{Introduction}

In the local Universe, there are strong correlations between galactic morphology, star formation rate, and stellar mass. The majority of low-mass galaxies are star-forming ``late-type'' disks (LTGs) while the most massive ones are quiescent ``early-type'' ellipticals or lenticulars \citep[ETGs; e.g.,][]{Schawinski2014}. The exceptions -- blue spheroids or red disks -- exist but are rare, less than a few percent of the overall population \citep[e.g.,][]{Masters2010,Barro2013}. While these correlations are well-established, their cause is still disputed: does the structure of a galaxy change after it quenches, or vice versa, and how does this depend on stellar mass?

\begin{figure*}
    \centering
    \includegraphics[width=\linewidth]{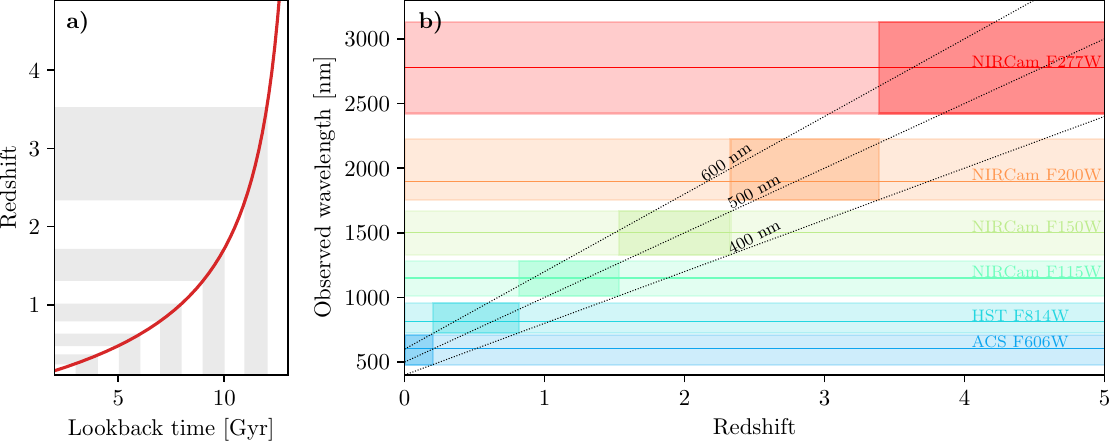}
    \caption{\textbf{a)} Redshift as a function of lookback time, showing our 1 Gyr bins that evenly sample lookback times from 2 to 13 Gyr. \textbf{b)} our choice of HST/JWST filters as a function of redshift to trace the rest-frame optical morphology. Each light region schematically represents the throughput and the pivot wavelength of each band. Dashed diagonals show the mapping of rest-frame 400, 500, and 600 nm to observed wavelengths. We assign bands to a redshift range such that the throughput is never shifted bluer than rest-frame 400 nm. This way, no flux is coming from blueward of the D4000 break, where galaxy SEDs change sharply depending on SFR, and we are always looking at intermediate stellar populations. Solid coloured regions show the redshift range where we use each band.}
    \label{fig:redshift}
\end{figure*}



To understand the co-evolution of galaxy star formation and structure, we must look at earlier epochs. At higher redshifts, the clean ``Hubble tuning fork'' changes: merging and irregular galaxies are more common \citep[e.g.,][]{Mortlock2013,Ferreira2022b}, compact star-forming galaxies appear at $z$$\sim$$2$ \citep[e.g.,][]{Barro2013}, and even more compact ``little red dots'' are now found at $z>2$ \citep[e.g.,][]{Matthee2024}. However, these galaxies are still relatively rare compared to the general population. \cite{Ferreira2023} and \cite{Kartaltepe2023} visually classified large samples of galaxies observed with JWST and found that a ``Hubble sequence'' as we know it persists for the majority of sources up to $z$$\sim$$5$. This result is still disputed, however: \cite{Pandya2024} find that galaxies visually identified as ``disks'' may instead be prolate, while several quantitative studies \citep[e.g.,][]{Shuntov2025} find that the early-type galaxies \new{classified as ETGs} lack concentrated bulges, challenging whether they are truly ``early-type''.\footnote{There is now a large body of work discussing the morphology of JWST-observed galaxies. References given here are just examples, trimmed for readability. We discuss the existing literature more fully in Sec. \ref{sec:discussion}}



One difficulty in studying structural evolution of galaxies is that identical sources look different at different distances. In addition to a simple change in brightness and angular size, we have to account for cosmological effects, as well as the intrinsic evolution in galaxy sizes and mass-to-light ratios \citep[e.g.,][]{vanderWel2014,Ren2024,McGrath2026}. Any approach to compare morphologies -- whether quantitative, machine learning-based, or visual -- has to account for this. For example, faint post-merger features are invisible at higher redshifts due to surface brightness dimming, and so neither computers nor scientists can detect them, resulting in a perceived lower fraction of tidal features at greater distances purely due to an observational bias \citep[e.g.,][]{Atkinson2013,Bilek2020}. Therefore, to \textit{robustly} compare galaxy structure across cosmic time, we must account for these observational biases. \new{The Hubble Space Telescope (HST) and the James Webb Space Telescope (JWST) are now providing high-quality observations of galaxies up to $z \sim 10$, allowing us to study morphological evolution of galaxies almost throughout their entire lifetime of the Universe -- but these biases must be carefully addressed.}


The goal of this work is to measure an ``absolute'' morphology of galaxies, as if observed at the same distance of 10 Mpc. This will result in consistent measurements across cosmic time, which will allow us to see the \textit{intrinsic} structural evolution, unaffected by observational biases. To do this, we projected images to the same absolute distance, matching the same effective resolution (\reseff{}) and surface brightness limit, accounting for cosmological effects. We then measured galaxy morphology using \texttt{statmorph-lsst}\footnote{\href{https://github.com/astro-nova/statmorph-lsst}{\texttt{statmorph-lsst}: github.com/astro-nova/statmorph-lsst}} \citep{statmorph,Sazonova2025} and applied dimensionality reduction to trace the evolution of galaxies in a morphological phase space. We used the publicly available multi-band data from the \textit{JWST} CEERS survey, which includes photometric redshifts, stellar mass, and star formation rates spanning $0.15<z<4.5$ range. This allowed us to connect the relative timelines of the evolution in galactic structure and star formation, and see whether one drives the other.

The paper is organized as follows: in Sec. \ref{sec:methods}, we present our sample selection, redshift matching procedure, and morphological analysis; in Sec. \ref{sec:umap} we analyze the morphological distribution of galaxies in $0.15<z<4.5$ range and in Sec. \ref{sec:evolution} we discuss its time evolution. Finally, we compare our findings to other works in Sec. \ref{sec:discussion}. Throughout the paper, we use the AB magnitude system and \cite{Planck2018} cosmology ($H_0$, $\Omega_m$) = (67.66, 0.3097).



\section{Methods}\label{sec:methods}

\subsection{Sample \& data selection}\label{sec:sample}

\begin{figure*}
    \centering
    \includegraphics[width=\linewidth]{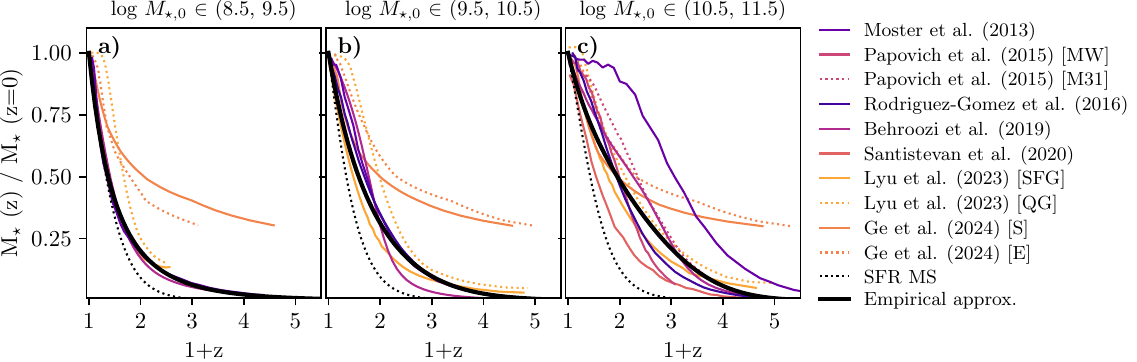}
    \caption{
    Stellar mass history for \textbf{a)} low, \textbf{b)} intermediate, and \textbf{c)} high-mass galaxies at $z=0$. We compiled mass history estimates from simulation-based and observational studies (colored lines) as well as from the star-forming main sequence \citep{Ormerod2024}, assuming purely in-situ growth (black dashed line). We use an approximation that averages between these studies (black solid line) in this work}
    \label{fig:mah}
\end{figure*}
We make use of the publicly available data from the JWST Cosmic Assembly Near-infrared Deep Extragalactic Legacy Survey \citep[CEERS;][]{ceers1,ceers2,ceers3} and the HST Cosmic Assembly Near-infrared Deep Extragalactic Legacy Survey \citep[CANDELS;][]{candels1,candels2}, together covering a large area of the sky in seven JWST and seven HST filters.

We selected our primary sample using the CEERS v1.0 data release \citep{ceersdr1} which is accompanied by a catalogue of photometric redshifts, stellar masses, star formation rates, and other galaxy properties obtained with spectral energy distribution (SED) fitting as described in \citep{Cox2025}. Since the goal of this project is to apply machine learning techniques to analyze the multidimensional morphology space, we required our sample to have a balanced representation of different galaxy masses and lookback times. We chose a sample that evenly spans three mass stellar mass bins with $\log M_\star = 8\sim11$, and 1 Gyr lookback time bins ranging as $t = 2 \sim 13$ Gyr. This means that while our sample is uniform in lookback time, it is uneven in redshift space (Fig. \ref{fig:redshift}a). For each mass and lookback time bin, we selected 100 galaxies from the CEERS catalog into our sample. Since there are relatively few low-redshift ($z<0.8$) galaxies in the CEERS field alone, we also included data and catalogs from the CANDELS COSMOS \citep{Nayyeri2017}, UDS \citep{Santini2015}, and GOODS-North \citep{Barro2019} fields for this bin.



Different galaxy components -- young and old stars, as well as gas -- are typically distributed differently in a galaxy, and so morphology changes across observational bands  \citep[e.g.,][]{Windhorst2002,Mager2018,Rutkowski2025}.
When analyzing morphology, it is thus essential to study the structure of the same component across the entire redshift range. A single band traces younger populations in more distant sources, which is known as a ``morphological k-correction'' \citep{Windhorst2002}. Therefore we made sure to choose a filter for each galaxy that best traces the same rest-frame wavelength. 

One prominent feature in a galaxy's SED is the 4000\r{A} break \citep{Hamilton1985,Balogh1999}: quiescent galaxies with older stellar populations have a sharp decline in energy blueward of 4000\r{A}, while star-forming galaxies have significant UV flux contribution from young stars. On the other hand, the SEDs of quiescent and star-forming galaxies in the optical are relatively flat. It is hence more important to ensure that our chosen filter never overlaps with the D4000 feature. Given each galaxy's redshift, we chose the \textit{bluest} possible filter such that its throughput does not extend beyond 4000\r{A}. We show our selection schematically in Fig. \ref{fig:redshift}b: the throughput of each filter is plotted as a light shading, with the solid lines indicating pivot (central) wavelengths. Grey dashed lines show the rest-frame 400, 500, and 600 nm as a function of redshift. The redshift range for which the filter is used is a darker shaded region. With our selection method, the central restframe wavelength of a filter ranges from 450 to 650 nm -- tracing intermediate-age stellar populations -- and we never look at the flux originating blueward of D4000 coming from young stars. This is especially important at higher redshifts, where star formation is more stochastic \citep[e.g.,][]{Clarke2024,Mitsuhashi2026}, and so the galaxy structure in the rest-frame UV can change on short timescales and is therefore not representative of broader evolution.

\subsection{Mass assembly history}

\begin{figure}
    \centering
    \includegraphics[width=\linewidth]{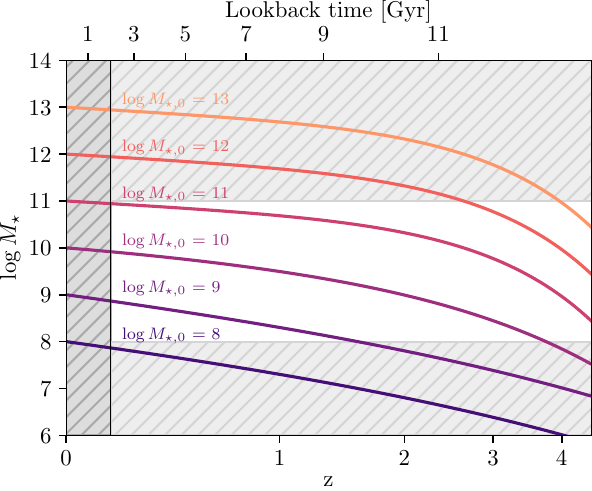}
    \caption{The mass assembly history of galaxies with different final mass, $M_{\star,0}$, using the curves in Fig. \ref{fig:mah}. Grey shaded regions show the redshift and stellar mass limits of our sample. Only the $9 < \log M_{\star,0} < 11$ region is relatively well-sampled in our lookback time bins.}
    \label{fig:completeness}
\end{figure}

The goal of this work was to look at the evolution of galaxy structure with cosmic time. By simply binning our galaxies into different mass bins, we can only compare properties of equal mass galaxies at different redshift. However, a $\log M_\star \sim 10$ galaxy at $z=2$ will accrete mass over time and should not be compared to a  $\log M_\star \sim 10$ at $z=0.15$. Instead, for the final analysis, we binned our sources into their expected $z=0$ mass, $\log M_{\star,0}$.

To do this we needed to assume a mass assembly history. We pooled the results from several studies, who investigated the mass assembly history for different $z=0$ stellar masses, both in observations and simulations. Fig. \ref{fig:mah} shows the mass assembly history from \cite{Moster2013}, \cite{Papovich2015}, \cite{RodriguezGomez2016}, \cite{Behroozi2019}, \cite{Santistevan2020}, \cite{Lyu2023}, and \cite{Ge2024}. We also added a dashed curve for an expected mass assembly history driven purely by star formation for a galaxy on a star-forming main sequence from \cite{Ormerod2024}, noting that this is an extremely simplistic approximation since it does not account for mergers or quenching events.

There is a significant scatter between different studies, even when binning the results in 1 dex mass ranges. In particular, the results of \cite{Ge2024} are different from all other studies. These are derived by modelling the star formation history of local galaxies, and the flattening at high $z$ is likely caused by the difficulty to distinguish very old stellar populations from one another \citep[e.g.,][]{CidFernandes2005}, so we do not consider this fit. At the high mass end, the \cite{RodriguezGomez2016} curve is much flatter at early times. It is likely because this curve is derived from the Illustris simulations where mass quenching is very rapid, and so we also do not include this curve.

\new{After removing the outlier assembly histories discussed above, we estimated the average assembly history in each of the three mass bins, giving the solid black curves shown in Figure \ref{fig:mah}. For each curve, we assumed the same function ansatz of the form:}

\begin{equation}
\new{
    M_\star (z) = M_{\star, 0} (1+z)^{-\alpha} e^{-(z/z_0)^\beta},
}\label{eq:mass_fun}
\end{equation}

\noindent where $M_{\star, 0}$ is the final stellar mass at the present epoch, and $\alpha$, $\beta$, and $z_0$ are variable parameters. For each final stellar mass bin, we approximated $\alpha$, $\beta$, and $z_0$ to roughly span the range of curves from previous studies. These values are listed in Table \ref{tab:mah}.

{\renewcommand{\arraystretch}{1.5}
\begin{table}
\caption{Mass accretion history parameters}
    \begin{tabularx}{\linewidth}{@{\extracolsep{\fill}}l l l l}
        \toprule
        \toprule
            $M_{\star, 0}$ bin & 
            $\alpha$ &
            $\beta$ &
            $z_0$\\
        \midrule
        (8, 9.5) & 1.6 & 1.0 & 2.0 \\
        (9.5, 10.5) & 1.2 & 1.6 & 2.0\\
        (10.5, 11.5+) & 0.9 & 2.5 & 2.5\\
    \bottomrule
    \vspace{-20pt}
    \end{tabularx}
    \label{tab:mah}
\end{table}
}

Then, for each galaxy in our sample, we placed it on the each of the three black curves at its observed redshift, and then determined an estimated z=0 mass based on each of the three curves. Finally we chose the $M_{\star,0}$ value closest to the midpoint of its bin. \new{While it is difficult to account for the stochasticity in the real star formation histories, in our analysis we estimated the uncertainties by bootstrapping the final stellar mass $M_{\star,0}$ from a Gaussian distribution $\mathcal{N} (M_{\star,0},~0.5~\rm{ dex})$.}



Our sample was chosen specifically to have a uniform sampling of stellar masses and lookback times, which means our sample is very non-uniform in terms of progenitor masses. Fig. \ref{fig:completeness} shows the average stellar mass of a galaxy as a function of redshift for progenitors of $\log M_{\star,0} \in (8, 13)$ galaxies. The horizontal hatched regions show stellar mass ranges excluded from our sample, and the vertical region shows our redshift cutoff. Given our lookback time bins, we only sample reliably the progenitors of galaxies in the $\log M_{\star,0} \in (9, 11)$ range. Outside this range we miss the low-mass progenitors at high redshifts and massive progenitors at low redshifts. Therefore, we focus the majority of our discussion on the two mass bins spanning this range.


\subsection{Preliminary image processing}
\begin{figure*}
    \centering
    \includegraphics[width=\linewidth]{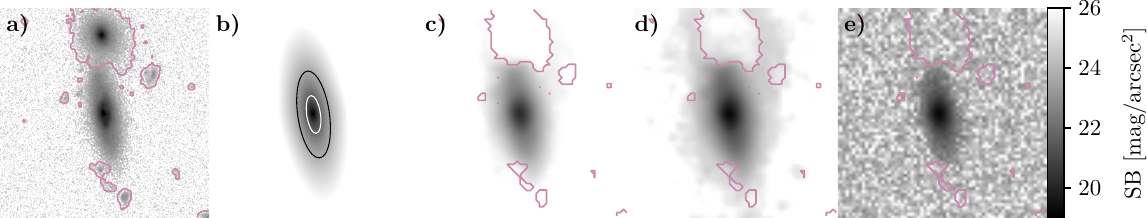}
    \caption{Example of the redshift matching procedure for one $z=0.556$ galaxy observed with HST F814W. \textbf{a)} A cutout of the CEERS field. We deblend and mask out any other sources in the cutout (purple contours). We then fit the galaxy with \textsc{Galfit} and \texttt{petrofit} (\textbf{b}) to derive the effective radii -- $R_{0.5}$ (white) and $R_p$ (black). Finally, we augment the image: convolve with a PSF to achieve an effective PSF FWHM of 3.8$R_p$, downsample the image such that $R_p$ spans 13.6 px (\textbf{c}), shift to an ``absolute'' magnitude system at 10 Mpc (\textbf{d}), and finally add noise to match a 1$\sigma$ surface brightness limit of 23.7 mag/arcsec$^2$ (\textbf{e}).}
    \label{fig:augments}
\end{figure*}

Using the parent CEERS and CANDELS catalogs, we constructed a preliminary sample of 3,000 randomly selected galaxies, uniformly spanning lookback time and stellar mass bins. We queried the Mikulski Archive for Space Telescopes\footnote{\href{https://archive.stsci.edu/}{MAST data archive: https://archive.stsci.edu}} (MAST) for archival imaging in the chosen filter for each source. We then inspected each image and removed any galaxies that were visibly artifacts, or were off the mosaic footprint for that particular filter, and replaced these with new galaxies from the parent catalog. 

For each galaxy, we made cutouts with a 30 kpc radius, which would span the full angular extent of even the largest, most nearby sources. We then ran a segmentation routine to detect the main source and mask out any foreground or background contaminants. A good segmentation map is crucial for any structural analysis, as a presence of contaminants will affect most quantitative parameters and produce a disturbed, merger-like measurement. For a true fair comparison, it may be beneficial to construct segmentation maps \textit{after} redshifting, since our ability to deblend sources depends on resolution and image depth. However, since nearby galaxies have a larger angular size, they are more likely to have projected background contaminants and so an accurate segmentation is more important. For this reason we opted to run segmentation first on the original images and mask out any extraneous sources before redshifting. 

We followed a ``hot-cold'' segmentation algorithm from \cite{Sazonova2025} to mask out potential bright foregrounds embedded into our targets. We briefly describe the algorithm and the important variables here and in Tab. \ref{tab:segmentation-params}, and refer the reader to \cite{Sazonova2025} and the \texttt{statmorph-lsst} package documentation for a more detailed description.

{\renewcommand{\arraystretch}{1.75}
\begin{table}
\caption{Segmentation parameters}
    \begin{tabular}{l l l}
        \toprule
        \toprule
            Param.  & 
            Description &
            Value  \\
        \midrule
        \texttt{hot\_thresh} & ``Hot'' SNR threshold & 5  \\
        \texttt{cold\_thresh} & ``Cold'' SNR threshold & 1.5 \\
        \texttt{min\_area} & Minimum segment area & 5~px \\
        \texttt{contrast} & Deblending contrast & 0.001  \\
        \texttt{overlap\_thresh} & \parbox{4.75cm}{Minimum overlap between a ``cold'' segment and a ``hot'' source} & 0.2 \\
    \bottomrule
    \end{tabular}
    \label{tab:segmentation-params}
\end{table}
}

\begin{enumerate}
    \item \textbf{Hot step:} detects all sources larger than \texttt{min\_area} above \texttt{hot\_snr} threshold above the standard deviation of the sky, $\sigma$. All ``hot'' sources other than the target are masked.
    \item \textbf{Cold step:} detects all sources larger than \texttt{min\_area} above the \texttt{cold\_snr} threshold. All ``cold'' sources other than the target are also masked
    \item \textbf{Deblending:} the final step aims to mask out any flux from ``hot'' foregrounds that contaminate the main source. We deblend the target aggressively (with a low \texttt{contrast} and large \texttt{nlevels}), and then mask any segments that have an overlap fraction with a masked hot segment above \texttt{overlap\_thresh}.
    \item \textbf{Grow mask:} finally, we grow each masked segment or source proportionally to its area to capture any extended faint flux.
\end{enumerate}

We manually examined the segmentation maps and changed the default parameters for a subset of galaxies where additional deblending was needed. The first panel of Fig. \ref{fig:augments} shows the masked sources for an example galaxy as pink contours.

Our goal was to match the effective resolution \reseff{}, the number of resolution elements spanning a galaxy, which required knowing the sources' radii. We measured S\'ersic half-light radii \citep[$R_{0.5}$;][]{Sersic1963} with \textsc{Galfit} \citep{galfit}, and Petrosian radii \citep[$R_p$;][]{Petrosian1976} as well as non-parametric half-light radii with \texttt{petrofit} \citep{petrofit} (Fig. \ref{fig:augments}, second panel). \new{These radii, along with other structural parameters we used in this work, are defined in Appendix \ref{app:params}.} We compared the S\'ersic and the non-parametric half-light radii and found that the S\'ersic radius is, on average, 1.74 pixels smaller because it accounts for the PSF smearing, but otherwise the agreement is good. Since not all galaxies had successful S\'ersic fits, we decided to use Petrosian radii. 

\subsection{``Absolute'' images}\label{sec:redshifting}

\begin{figure*}
    \centering
    \includegraphics[width=\linewidth]{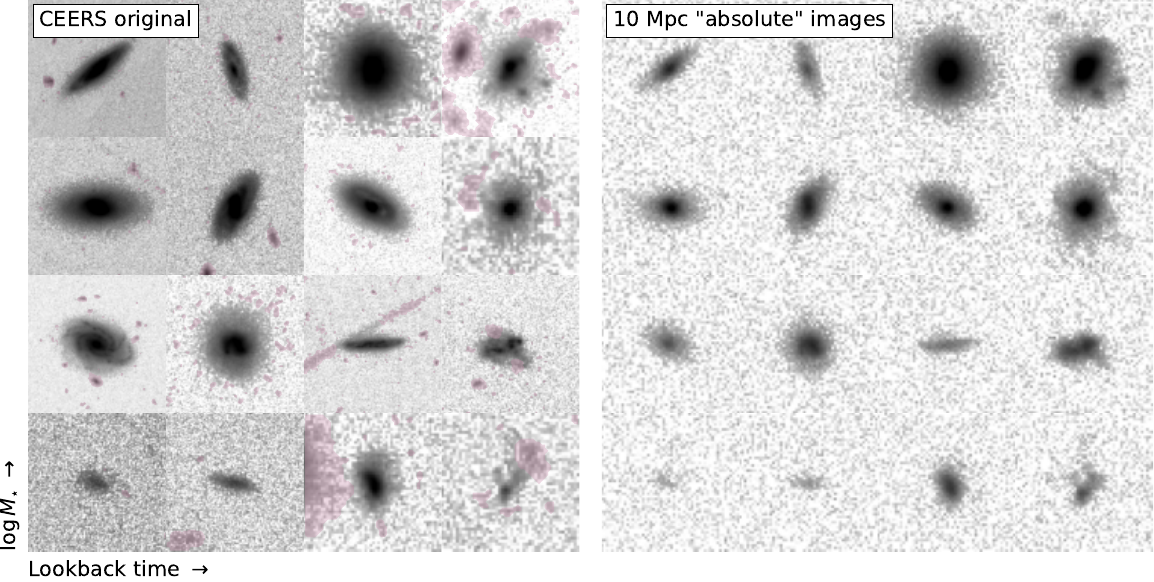}
    \caption{The original HST/JWST images (\textbf{left}) and our augmented ``absolute'' images (\textbf{right}) for a subset of galaxies in our sample. Purple filled contours show masked and noise-filled sources. Each image spans 2$R_p$ of the galaxy. The apparent decrease in the intrinsic size is a consequence of adding noise to match a uniform surface brightness limit, which affects the dim features of low-mass galaxies more.}
    \label{fig:augments_comp}
\end{figure*}

Since morphological parameters depend heavily on resolution, signal-to-noise, or both \citep[][and references therein]{Sazonova2025}, we opted to compute structural measurements on ``absolute''\footnote{We borrow the terminology from absolute magnitudes, originally intended to measure brightness of stars as if they are 10 pc away. The angular scale would be obviously ridiculous if we placed our galaxies at 10 pc, so we move them to a more realistic ``absolute'' distance of 10 Mpc.} images with a matched resolution and depth, as if the galaxies are observed at the same distance of 10 Mpc. There are, however, several caveats to achieving this in a way such that morphological measurements are truly consistent between galaxies of different masses and at different redshifts.

A simple choice would be to ensure the same absolute surface brightness limit (\sblim{}) and the same physical resolution in parsecs per pixel, and we explore this choice in Appendix \ref{app:simple_augments}. However, \cite{Sazonova2025} showed that the bias in most morphological parameters depends on the \textit{effective} resolution rather than the absolute. At a fixed physical resolution, the \reseff{} is lower both for lower-mass galaxies due to the mass-size relation \citep[e.g.,][]{Shen2003,vanderWel2014}, and for more distant galaxies due to the size evolution at fixed mass \citep[e.g.,][]{vanderWel2014,Allen2024,McGrath2026}. On the image depth side, matching \sblim{} is a good choice since it allows us to trace a consistent light (and by proxy, mass) density; but at higher redshifts the optical mass-to-light ratio changes as star formation surface density increases \citep[e.g.,][]{Conselice2005,Bezanson2013,Sande2015,Yu2023}. This leads to an optical luminosity evolution at fixed mass, and while the \textit{light} density is consistent across redshifts at fixed \sblim{}, the \textit{mass} density is not. The approach we took in this work is to account for all of these effects -- cosmological surface brightness dimming, angular size change, as well as galaxies' intrinsic size and luminosity evolution. 

First we matched the effective resolution. 90\% of our sample have $R_p$ larger than 13.6 pixels, and a PSF FWHM smaller than 0.26$R_p$, so we chose these limits. We convolved each image with a Gaussian PSF such that the effective FWHM of the empirical JWST PSF convolved with a Gaussian is 0.26$R_p$, and then resized the image with \texttt{reproject} \citep{reproject} such that $R_p$ spans 13.6 pixels, \new{effectively giving us 4 PSF resolution elements per galactic radius} (Fig. \ref{fig:augments}c). We also downsampled the convolved PSF and saved it to use in analysis. This resampling accounts for any mass- or redshift-dependent size evolution automatically, since we rescaled each galaxy based on its own intrinsic size, and $R_p$ is robust to changes in resolution and depth \citep{Lotz2004,Sazonova2025}. 

Next we had to match the image depth. \cite{Yu2023} studied the intrinsic luminosity evolution of galaxies at fixed different morphological type and mass. In general, luminosity evolution has the form $L(z) = L_0 (1+z)^{\alpha}$, where $\alpha$ depends on the subsample. At rest-frame 500 nm, $\alpha$ is roughly 1.0 for the entire mass range for both late- and early-type galaxies, so we opted to use a single value $\alpha = 1.0$ for our entire sample. To account for the evolution in mass-to-light ratio, we divided the flux of each image by $(1+z)$ to recover an expected $L_0$ luminosity. We then converted the image into ``absolute'' magnitude units, placing the source at 10 Mpc, and accounting for the AB magnitude zeropoint change as we moved to a bluer filter at rest-frame (Fig. \ref{fig:augments}d). Effectively this means that the surface brightness limit decreases by $7.5 \log (1+z)$ \citep{Oke1968,Hogg2002}. 

Finally, we calculated a sigma-clipped standard deviation of the background on these absolute images to compute 1$\sigma$ \sblim{}. 90\% of the sample have \sblim{} deeper than 23.7 mag/arcsec$^2$, so we added Gaussian white noise to achieve this limit to the images. The amount of noise to add, $\sigma'$, is given by 

\begin{equation}
    (\sigma')^2 = \sigma^2 \left( 10^{2\left(\sblimm^{\rm{new}} - \sblimm^{\rm{og}}\right)/2.5} - 1 \right).
    \label{eq:main_sequence}
\end{equation}

\noindent where $\sigma^{\rm{og}}$ and $\sblimm^{\rm{og}}$ are the original noise level and surface brightness limit, and the desired new surface brightness limit is $\sblimm^{\rm{new}}=23.7$ mag/arcsec$^2$. Adding the noise gave us the final augmented image (Fig. \ref{fig:augments}e). Figure \ref{fig:augments_comp} shows thumbnails of 16 galaxies from our sample, spanning a range of redshifts and stellar masses, before and after our image matching procedure.  Galaxies in our ``absolute'' images have similar image quality across our redshift range, which lets us compare morphologies without worrying about resolution or depth-induced biases. Our final sample has 2825 galaxies.

\subsection{Morphological analysis}

We computed structural parameters of the galaxies in our sample on the absolute images with \texttt{statmorph-lsst}. The measurements can be split roughly into three groups: geometric measurements (centroids, radii, ellipticity, orientation), bulge strength measurements (concentration, Gini, $M_{20}$, and derived values), and disturbance measurements (asymmetry, Gini-$M_{20}$ disturbance, smoothness, substructure, and isophotal asymmetry). Finally, we defined two additional concentration-like metrics -- the ratios of 20\ts{th} to 50\ts{th} isophotes similar to that of \cite{Trujillo2001} and \cite{Graham2001}, as well as the ratio of 80\ts{th} to 50\ts{th} isophotes. \new{Appendix \ref{app:params} lists all the parameters available in \texttt{statmorph-lsst}, their brief descriptions, and notes the ones we used in this analysis. Since we are interested purely in structural evolution, we did not use any of the radius or flux measurements.} We additionally performed S\'ersic fits with \textsc{Galfit}; however, since the fits failed for 10\% of our sample, we did not include S\'ersic parameters in our analysis.  We discuss the relationship between S\'ersic indices and non-parametric morphology in Sec. \ref{sec:umap} and show that our results do not change by including the S\'ersic index in Appendix \ref{app:sersic_umap}. Overall, we computed 59 parameters and used 24 of them in our analysis, as outlined in Table \ref{tab:parameters}.

This is a relatively large number of features, and it is challenging to interpret them in tandem. We show the distributions of these parameters as a function of redshift and mass in Appendix \ref{app:parameter_dist}, but for the main body of this work, we turned to a machine-learning driven approach. 

\subsection{UMAP analysis}

To interpret this feature space, we used the Uniform Manifold Approximation \& Projection \citep[UMAP;][]{umap} -- a dimensionality reduction technique. A UMAP constructs a fuzzy nearest-neighbors graph in the $N$-dimensional feature space, and then finds the most similar lower-dimensional graph, preserving both global and local structure of the connections. We chose to reduce our dataset to two dimensions, $U_1$ and $U_2$. Dimensionality reduction techniques such as the principal component analysis (PCA) are common in astronomy \citep[e.g.,][]{Conselice2006}, but PCA assumes the data are linearly distributed. UMAP is much more flexible due to its graph-based approach and can capture non-linear relationships between the data points in the feature space. UMAP is commonly used to visualize the feature space of large neural networks in two dimensions \citep[e.g.,][]{Tohill2024,Desmons2024}, including for a morphological analysis and classification \citep{Fang2026} but is now being increasingly used as a standalone technique in analysis as well \citep[e.g.,][]{Rosito2023,Haggar2024,Narayan2025}. 

We transformed our dataset of 2825 galaxies with 24 parameters each into a two-dimensional $(U_1, U_2)$ space. We used UMAP parameters optimized for global clustering, with a large number of neighbors and sampling rate, which ensures stability of the overall structure. We chose \texttt{n\_neighbors}, the number of neighbors to consider during fitting, to be 100, which prioritizes global structures. We also set \texttt{negative\_sample\_rate} and \texttt{transform\_queue\_size}, parameters that govern how many points are considered simultaneously during fitting, to 100, since higher values increase accuracy and stability of the manifold at a greater computational cost.

\section{Results} \label{sec:results}

\subsection{The morphological distribution of galaxies \& the Hubble Sequence}\label{sec:umap}

\begin{figure*}[p]
    \centering
    \includegraphics[width=\linewidth]{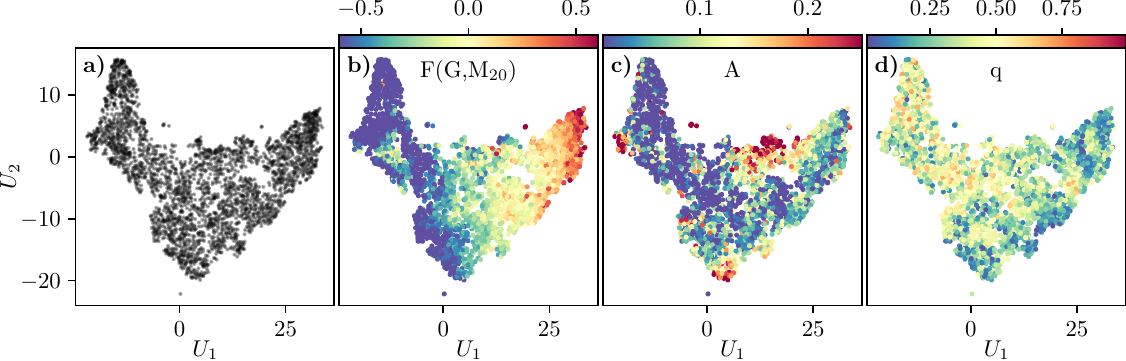}
    \caption{(\textbf{a}) The distribution of the entire sample in the UMAP space ($U_1$, $U_2$). Each dot is a galaxy. Panels \textbf{b}), \textbf{c}), and \textbf{d}) show how three parameters -- Gini-M$_{20}$ bulge strength, asymmetry, and axis ratio -- change across the UMAP space. $U_1$ globally maps the Hubble sequence, with late-type galaxies on the left and early-type on the right. Galaxies on the edges of the $U_2$ distribution are more disturbed with higher asymmetry. The distributions of the remaining parameters are in Appendix \ref{app:parameter_dist}.}
    \label{fig:umap}
\end{figure*}

Fig. \ref{fig:umap}a shows the distribution of our dataset in the UMAP plane, where each dot represents a galaxy. Panels b), c), and d) show the same distribution, this time coloring each point by the Gini-$M_{20}$ bulge strength ($F(G,M_{20})$; Fig. \ref{fig:umap}b), asymmetry ($A$; Fig. \ref{fig:umap}c) and ellipticity ($e$; Fig. \ref{fig:umap}d) respectively. These three parameters are a subset of the 24 metrics seen by the UMAP that we chose as an example -- we show the equivalent plots using the remaining parameters in Appendix \ref{app:parameter_dist}.


The distribution of $(U_1, U_2)$ values in Fig. \ref{fig:umap}a is clearly a continuous single cluster. UMAP is a powerful clustering tool, and if there existed distinct classes of objects in the large feature space, the UMAP projection can easily help find them in 2D \citep[e.g., as was done in a different morphological analysis in][]{Fang2026}. The fact that our data lies on a single continuum then implies that there are no distinct groups of galaxies based on their structure. Instead of a bimodality we see a single sequence of galaxy morphology, coincidentally resembling a ``U'' shape (or a bat!).

\begin{figure*}[p]
    \centering
    \includegraphics[width=\linewidth]{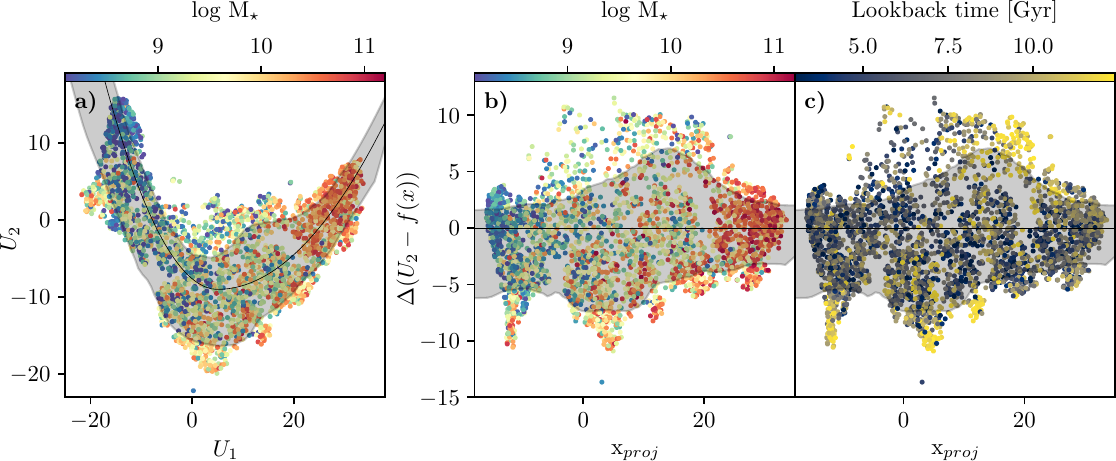}
    \caption{(\textbf{a}) The distribution of stellar mass across the UMAP space. Stellar mass increases along a parabolic curve defined in Eq. \ref{eq:mass_fun} (solid black line). Panels \textbf{b)} and \textbf{c)} show the same data, now projected onto the curve. The grey shaded region spans the 16\ts{th}/84\ts{th} quantiles of the distribution. We define this region as a mass-morphology ``main sequence'. As seen in panel \textbf{c)}, galaxies in the ``irregular'' region are preferentially found at higher lookback times.}
    \label{fig:umap_mass}
\end{figure*}

\begin{figure*}[p]
    \centering
    \includegraphics[width=\linewidth]{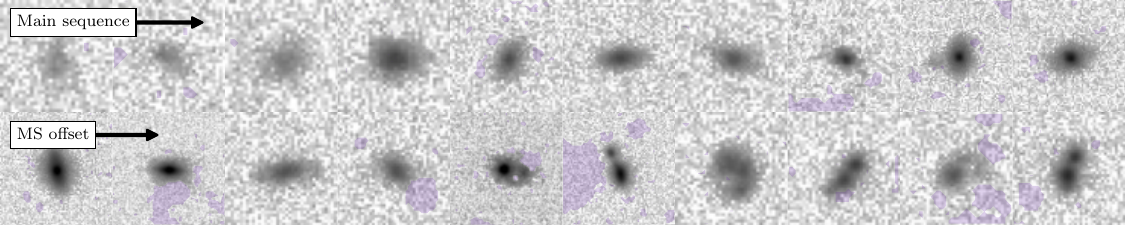}
    \caption{Examples of galaxies along the ``main sequence'' (top), sorted by $x_{\rm{proj}}$, and outside the ``main sequence'' (bottom), sorted by their distance away from the ``main sequence''. The top row shows a clear Hubble-like sequence from late- to early-type galaxies while the bottom row contains more disturbed galaxies and clear mergers. Violet regions show masked and noise-filled pixels.}
    \label{fig:main_sequence}
\end{figure*}

\clearpage

\begin{figure*}[t]
    \centering
    \includegraphics[width=\textwidth]{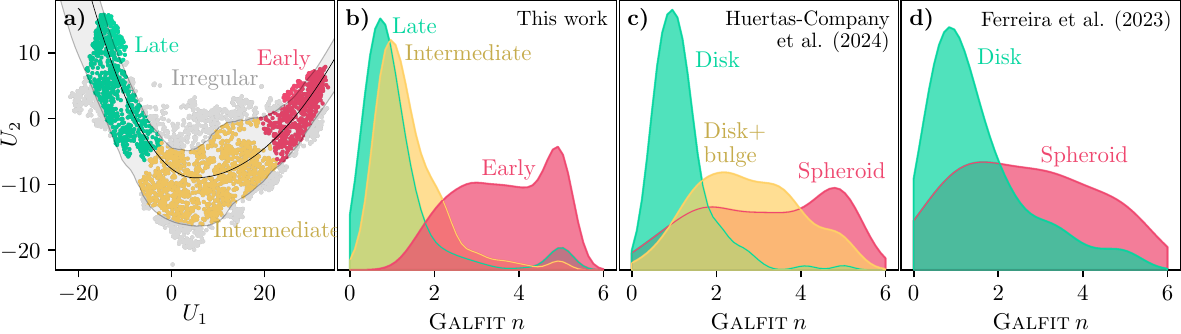}
    \caption{\textbf{a)} Our classification of the UMAP phase space into morphological types -- ``late'' (green), ``early'' (red), ``intermediate'' (red) and ``irregular'' (grey).  \textbf{b)} The distribution of S\'ersic indices for these categories. ``Late'' and ``intermediate'' groups have $n<2$ and we collectively refer to them as Late-Type Galaxies (LTGs), while the ``early'' group has $n>2$, and so we call them Early-Type Galaxies (ETGs). \textbf{c)} and \textbf{d)}: the distribution of S\'ersic indices for \cite{HuertasCompany2024} deep learning and \cite{Ferreira2023} visual classifications. Our parametric approach gives a better split between late- and early-type galaxies than visual classifications, but comparable to deep learning. }
    \label{fig:morph_types}
\end{figure*}

When we color the points by the measured parameters -- bulge strength, asymmetry, and axis ratio (Fig. \ref{fig:umap}b,c,d) -- we can start to interpret this sequence. The upper-left side hosts galaxies with low bulge strength, low asymmetry, and typically higher ellipticities: these are characteristic of ``late-type'' morphologies. The upper right side, on the other hand, has high bulge strength and low to intermediate asymmetry -- these are typical ``early-type'' galaxies\footnote{While we expect LTGs to have higher intrinsic asymmetries due to star-forming knots, ETGs are more massive and hence have higher \avgsnr{} in our images. Since asymmetry is very sensitive to \avgsnr{}, our ETG population appears more asymmetric, on average.}. The overall body of the shape represents a smooth transition between late- and early-type structure, correlating mostly with $U_1$. This is the global structure of the UMAP.

The local structure is more intricate, and there are regions of the UMAP with distinct morphological characteristics. \new{There are subtle gaps separating the left and the right ``wings of the bat'' from its ``body''. There are also distinct clusters:} the leftmost cluster hosts galaxies with low bulge strength but high asymmetry, while the upper and the lower middle ridges represent galaxies with intermediate- to high bulge strengths and high asymmetry. In general, the disturbed objects are found on the edges of the distribution, ether above or below the ``main sequence'' along the ``U'' shape.

To better understand the distribution, we then colored the points by stellar mass rather than a structural measure (Fig. \ref{fig:umap_mass}a). There is a clear mass trend, with low- and high-mass galaxies residing on the left-most and right-most sides of the U, respectively. This is a remarkable correlation, considering stellar mass was not a parameter known to the UMAP, and it reflects the link between stellar mass and morphology that appears to be present at all cosmic times in our sample.

We used Fig. \ref{fig:umap_mass}a to approximate a mass-morphology ``main sequence'' in the UMAP plane, which we represented as a piecewise quadratic to capture the asymmetry between the left and the right arms of the ``U''. The functional form of this main sequence, estimated empirically, is:

\begin{equation}
U_2 = \lambda (U_1 - 5) - 9,\; \qquad
\lambda=\begin{cases}
0.055 & U_1<5\\
0.02 & U_1\ge5
\end{cases}.
\end{equation}

Fig. \ref{fig:umap_mass}b shows the projection of the sample onto the ``main sequence'', and the grey shaded region encompasses the running 16\ts{th}/84\ts{th} quantiles of the offset from the ``main sequence''. Fig. \ref{fig:umap_mass}c shows the same distribution, this time colored by lookback time. \new{Example galaxies along this ``main sequence'' and the outliers are plotted in Fig. \ref{fig:main_sequence}}. The ``main sequence'' captures the transition from low-mass, late-type galaxies to massive early-type ones -- a continuum akin to the Hubble sequence we see in the local Universe. There does not seem to be an age gradient along the main sequence, so our results suggest that this Hubble sequence and a mass-morphology relationship exists at all cosmic times and is established early, agreeing with previous results, which we discuss in detail in Section \ref{sec:discussion}. On the other hand, the picture is different off the ``main sequence'' - galaxies that are ``outliers'' have typically high lookback times, hinting that mergers and irregular structures may be more common at higher redshifts, which we investigate further in Sec. \ref{sec:evolution}.

\subsubsection{Morphological classifications}\label{sec:classifications}

\new{While the UMAP generally shows a continuous distribution of galaxies instead of distinct classes, to aid the discussion, we split the UMAP space into four broad classes: ``late'', ``intermediate'', `early'', and ``irregular'' regions.}

\new{We used the UMAP shape and the distance along $x_{\rm{proj}}$ to separate galaxies on the ``main sequence'' into the three sub-types. Since there are subtle gaps in the distribution around $x_{\rm{proj}}=-3.5$ and $x_{\rm{proj}}=21$, this is where we drew the distinction, noting that this is an arbitrary choice. } We also included some galaxies off the ``main-sequence'' into ``late'' and ``early'' classes, as discussed below in Sec. \ref{sec:irregulars}. Fig. \ref{fig:morph_types}a) shows our final morphological classification with ``late'' galaxies in green, ``intermediate'' in yellow, ``early'' in red, and ``irregular'' in grey.


\new{Fig. \ref{fig:morph_types}b) shows the distribution of S\'ersic indices for the three types on the mass-morphology main sequence. The ``early'' group has distinctly high S\'ersic indices characteristic of early-type galaxies, with median $\langle n \rangle = 3.7\pm1.3$, where the uncertainties are the 1$\sigma$ quantiles of the distribution. The ``late'' and ``intermediate'' group have consistent S\'ersic indices, with $\langle n \rangle=0.8_{-0.4}^{+0.7}$ and $\langle n \rangle=1.2_{-0.5}^{+0.9}$. The main difference between these two populations is the stellar mass and hence \avgsnr{}, but they are otherwise morphologically similar. The difference between ``early'' and ``intermediate'' groups is, on the other hand, pronounced despite our rough choice of the $x_{\rm{proj}}$ boundary. For the remainder of the discussion, we refer to the ``early'' class as ETGs, and to the ``late'' and ``intermediate'' classes as LTGs.}

\new{For comparison, we also show the distribution of S\'ersic indices measured on our ``absolute'' images for a subset of galaxies with deep learning classifications from \cite{HuertasCompany2024} and visual classifications from \cite{Ferreira2023}. The deep learning classification defined three distinct classes, well-separated in their S\'ersic index, with $\langle n \rangle=3.2_{-1.9}^{+1.8}$ for spheroids, $\langle n \rangle=2.6_{-1.1}^{+1.4}$ for mixed bulge+disk systems, and $\langle n \rangle=1.0_{-0.4}^{+0.5}$ for disks. On the other hand, there is significant overlap in the ``spheroid'' and ``disk'' populations of \cite{Ferreira2023}, with $\langle n \rangle=2.4_{-1.3}^{+2.3}$ and $\langle n \rangle=1.2_{-0.6}^{+1.4}$, respectively. We discuss the differences between visual and parametric classifications along the Hubble sequence in more detail in Sec. \ref{sec:discussion}.}



\subsubsection{The Irregulars}\label{sec:irregulars}

\begin{figure*}[!h]
    \centering
    \includegraphics[width=\linewidth]{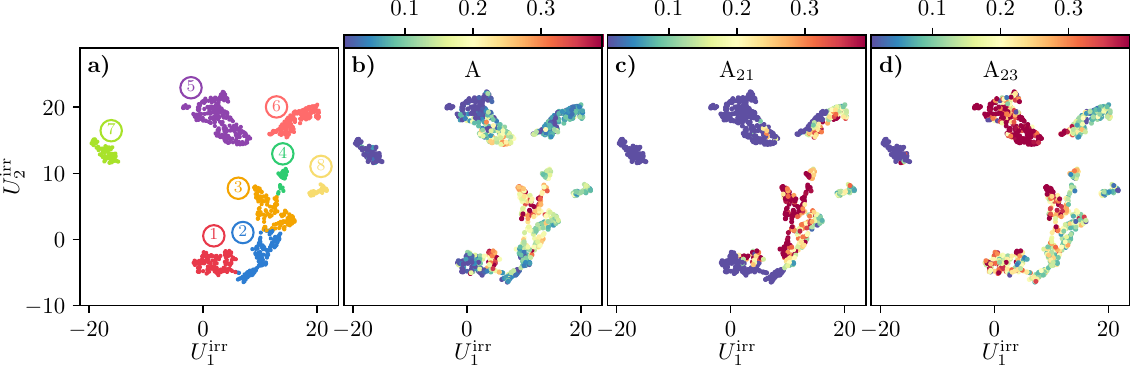}
    \caption{Same as Fig. \ref{fig:umap}, for a UMAP ran only on ``main sequence'' outlier galaxies. This created a new distribution of $(U_1^{\rm{irr}},U_2^{\rm{irr}})$, with distinct clusters, or sub-types of outlier galaxies. We used spectral clustering to separate the data into groups (\textbf{a}), and show the distribution of asymmetry / isophotal asymmetry in these groups in panels \textbf{b}, \textbf{c}, and \textbf{d}.}
    \label{fig:umap2}
\end{figure*}

The  outliers in the UMAP space are a heterogeneous population. Some galaxies may have regular shapes that happened to be outside an arbitrarily chosen 1$\sigma$ quantile range, while other galaxies may be galaxies with disturbed morphologies or mergers of different stages and mass ratios. To investigate this further, we excluded all ``main sequence'' galaxies and ran the UMAP algorithm again, to generate a new phase space of the ``main sequence'' outliers. This second distribution is shown in Fig. \ref{fig:umap2}.

The second run of the UMAP produced new phase space coordinates, $U_1^{\rm{irr}}$ and $U_2^{\rm{irr}}$. The distribution is much less continuous, with a few clear clusters, indicating that there are several distinct types of galaxies in the outlier population. We assigned galaxies to eight different clusters by running \textsc{Scikit-Learn} \texttt{SpectralClustering} -- these labels are shown as different colors in Fig. \ref{fig:umap2}a.

Panels b), c), and d) then color the points by three different asymmetry metrics -- CAS, 21 mag/arcsec$^2$, and 23 mag/arcsec$^2$ isophotal asymmetries respectively. Different groups in the phase space have different characteristic disturbances. For example, groups 1, 2, 3, and 4 have high $A_{21}$, i.e. bright asymmetric features such as double nuclei, while groups 5 and 6 have low $A_{21}$ but high $A_{23}$, indicating lower surface-brightness disturbances such as tidal tails. Group 7 has low asymmetry values at all levels.

\begin{figure*}
    \centering
    \includegraphics[width=\linewidth]{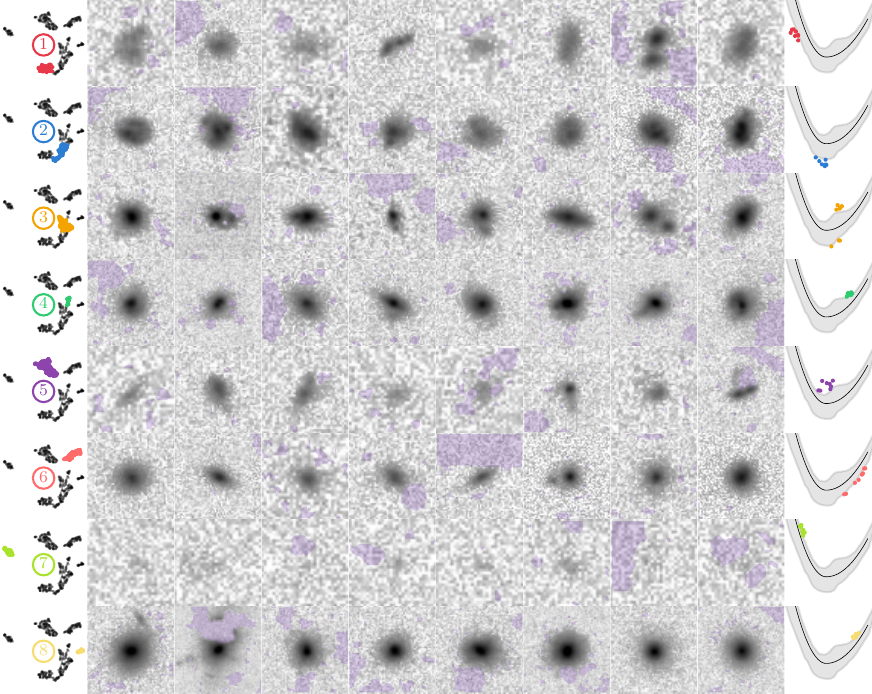}
    \caption{Example thumbnails of galaxies from each cluster in Fig. \ref{fig:umap2}. The locations of these clusters in the new $(U_1^{\rm{irr}},U_2^{\rm{irr}})$ and the original $(U_1, U_2)$ spaces are shown in the left-most and the right-most panels. While groups 1--6 show disturbed structures, group 7 and 8 galaxies do not, so we assign them to the ``late'' and ``early'' type groups, respectively.}
    \label{fig:irregulars}
\end{figure*}

To further interpret the meaning behind these clusters, we show example galaxies from each group in Fig. \ref{fig:irregulars}, where each row corresponds to a different cluster. The thumbnails on the left and right show where these galaxies reside in the outlier and the original UMAP spaces respectively. 

Group 7 galaxies are clearly galaxies with low \avgsnr -- these are $\log M_\odot \sim 8$ galaxies whose surface brightness is too low for us to measure their morphology in sufficient detail. These galaxies are similar to the late-type galaxies in Fig. \ref{fig:main_sequence}, except with particularly low surface brightness. We have no way to tell whether they are disturbed -- while their low surface brightness is unusual enough to make them 1$\sigma$ outliers, we consider these galaxies as ``late-type'', not irregular, in \new{Fig. \ref{fig:morph_types} and} our further analysis.

Similarly, Group 8 galaxies appear to be particularly concentrated early-type galaxies without any visually obvious disturbances. Their bright nuclei make them outliers from the typical early-type population of Fig. \ref{fig:umap_mass}, and this population might be just an extreme case of early-type galaxies. \new{Our main sequence is defined via quantiles, and so it is natural that some galaxies will lie outside those contours if they have extremely concentrated morphologies.} These galaxies have moderate $A$ and $A_{21}$ indices (Fig. \ref{fig:umap2}) indicating there may be some internal disturbances. However, since visually they look like regular galaxies, and they lie on the ridge of the ``early-type'' sequence in Fig. \ref{fig:umap_mass}, we consider these ``early-type main sequence'' galaxies in the future analysis.
 
The remaining groups exhibit visually obvious disturbances and so we classify all the galaxies in groups other than 7 and 8 as disturbed.

The large \new{contiguous} region, split into groups 1, 2, 3, and 4, appears to roughly map out a merger sequence among high surface-brightness, massive galaxies. \new{Group 3 mainly hosts very compact, bright galaxies with high $A_{21}$, i.e. high surface brightness disturbances, and several objects with double nuclei -- these are most typical of post-merger morphologies. Group 2 galaxies also have a disturbed internal structure but at lower surface brightness levels, with asymmetric envelopes, clumps, and an overall irregular shapes -- appearing similar to late-stage mergers or very clumpy galaxies. Group 4 galaxies have more tidal features than other groups, potentially suggesting a post-coalescence stage.} Finally, group 1 contains many obvious nuclei similar to group 3. Group 1 in particular is an interesting outlier group in the original ($U_1$, $U_2$) space. These galaxies have low concentration indices due to the two nuclei biasing the concentration measurement, and so reside in the ``late-type'' part of the UMAP despite being primarily made of massive, bright, and spheroidal sources. 

\new{Groups 5 and 6 are the final two distinct groups, mainly hosting lower surface brightness galaxies that still have disturbed features. Group 5 has high $A_{23}$, indicating low-mass galaxies with tidal features or asymmetric disks, whereas group 6 is similar to the 1--4 region but in lower-mass systems.}

\new{Overall, the ``irregular'' population clearly contains several distinct classes of objects, and the UMAP approach is a promising way to distinguish between them. With the available data, we can only speculate that these classes correspond to distinct stages of a merger sequence. However, \cite{Snyder2015a} demonstrated that morphological parameters derived from Illustris mock merger simulations evolve systematically with merger stage, suggesting they can constrain the merger phase when used in combination. In a future study, we will apply the UMAP approach to simulated galaxies with known merger stages to test whether distinct regions of UMAP space genuinely trace distinct merger phases.}

\subsection{Structural evolution over time}\label{sec:evolution}

The projected $z=0$ masses allow us to split our sample into the expected mass bins at the present time, and hence look at the evolution of the \textit{progenitor} of modern-day galaxies. 

\begin{figure*}
    \centering
    \includegraphics[width=\linewidth]{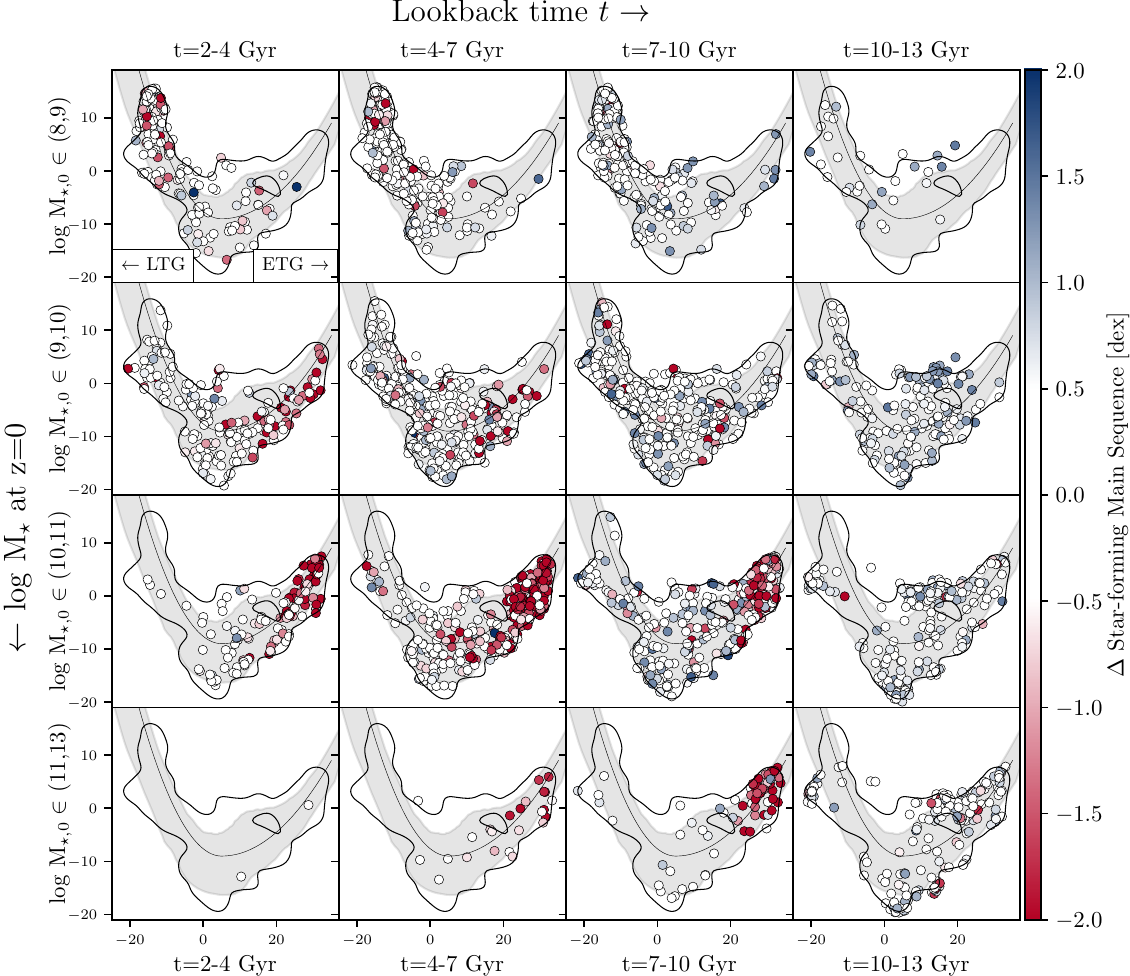}
    \caption{Evolution of galaxies in the UMAP plane at different lookback times. Each row represents a $z=0$ mass bin, from lowest- to highest-mass descendants. The color shows the star-forming main sequence offset. The contour outlines the full UMAP space (Fig. \ref{fig:umap}), the solid parabola and the grey shaded regions are the morphological main sequence from Fig. \ref{fig:umap_mass}. \textbf{High-mass end} (bottom row): progenitors of massive galaxies start off in the early-type/disturbed region while still star-forming, then quickly quench, retaining their morphology. \textbf{Milky-Way mass}: similarly, these start with either an early-type of disturbed morphology. ETGs then quench, while the disturbed galaxies continue star-forming. Some later move towards the late-type region. \textbf{Low-mass end} (top rows): low-mass progenitors exhibit a mix of late-type and disturbed morphologies at high $z$. For the most part, their morphology does not change over time.}
    \label{fig:evolution}
\end{figure*}

In Fig.  \ref{fig:evolution}, we show how the distribution of galactic morphologies changes for progenitors of different $\log M_{\star,0}$ galaxies (rows) at different cosmic times (columns). We colored each point by its offset from the star-forming main sequence from \cite{Ormerod2024} at its redshift, so that red points are quiescent, and blue are starbursts.  The grey shaded region shows the mass-morphology ``main sequence'' from Fig. \ref{fig:umap_mass}.

First, the consequence of our completeness limits shown in Fig. \ref{fig:completeness} are also clearly seen here: we are missing a large fraction of high-redshift low-mass and low-redshift massive progenitor galaxies. 

Qualitatively looking at the lowest- and the highest-mass bins, the distribution of morphologies in the UMAP space does not change significantly over time. Progenitors of low-mass galaxies always reside primarily in the ``late'' region, while the progenitors of massive galaxies are primarily in the ``early'' region. We can therefore conclude that, in the phase space probed by our UMAP, galaxy structure in these mass regimes \textit{does not change}. On the other hand, the star formation properties do change, as we see a larger fraction of quiescent galaxies at later times in both these groups. 

To explore these results further, we look at the fraction of galaxies of different morphological types and the fraction of quiescent galaxies in two different mass bins. Due to our completeness limits, we focus on two mass bins: $\log M_{\star,0} \in (9, 10)$ and $\log M_{\star,0} \in (10, 11)$.

\subsubsection{Progenitors of low-mass galaxies}

\begin{figure*}
    \centering
    \includegraphics[width=\linewidth]{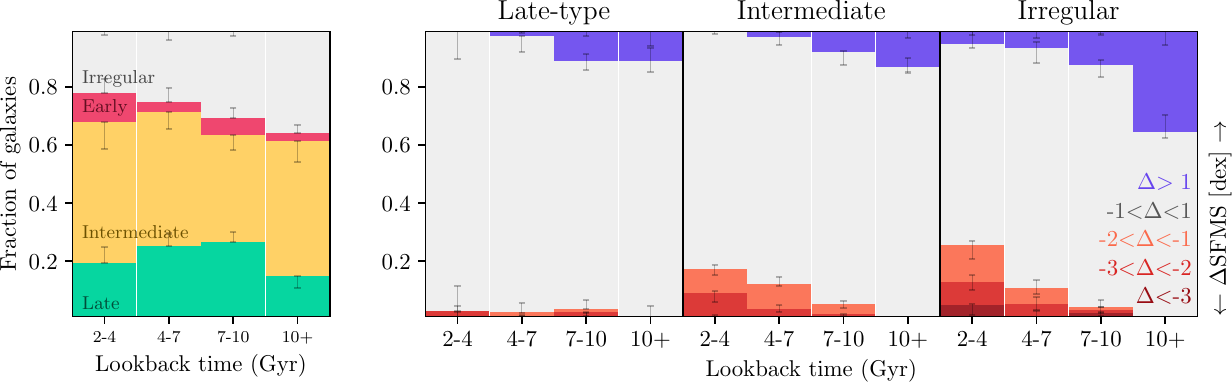}
    \caption{\textbf{a)} The fraction of ``late'' (green), ``intermediate'' (yellow), ``early'' (red), and ``irregular'' (grey) galaxies in the $\log M_{\star,0} \in (9,10)$ mass bin as a function of lookback time. \new{The error bars are obtained by bootstrapping $M_{\star,0}$ with a 0.5 dex uncertainty.} We consider both ``late'' and ``intermediate'' groups as LTGs. Galaxies in this final mass range have primarily late-type morphology at all cosmic times, with a constant $\sim$25\% irregular fraction. \textbf{b)} The fraction of galaxies in each morphological type that are within 1 dex of SFMS (blue), on the SFMS (grey), or are off the SFMS (each darker shade of red corresponds to a 1 dex offset). Low-mass progenitors are primarily star-forming at all cosmic times, with a slight increase in quenched fraction at later epochs.}
    \label{fig:qfrac_lowm}
\end{figure*}

Fig. \ref{fig:qfrac_lowm}a shows the fraction of ``early''-type, ``late''-type, intermediate, and irregular galaxies in the $9 < \log M_{\star,0} < 10$ mass bin as a function of lookback time. \new{The error bars reflect bootstrapping the final mass from a $\mathcal{N}(M_{\star,0},~0.5~\rm{ dex})$ distribution.} We focus on this mass bin rather than the lower one because our lowest progenitor mass-bin is not complete at high $z$ (Fig. \ref{fig:completeness}), we cannot fully sample progenitors on the lowest-mass end. 

In this mass bin, the dominant population is ``late'' and ``intermediate'' groups at all lookback times, which we both assigned as LTGs (Sec. \ref{sec:umap}). \new{The main difference between the two is that the ``late'' group is comprised of lower surface brightness, marginally detected disks, and the ``intermediate'' group galaxies have typical disk-like morphologies.} Combining these groups, we see that the overall LTG fraction is constant with redshift, at $\sim 65\%$. This agrees with other studies \citep[e.g.,][]{Lee2022,Ferreira2022a,Ferreira2022b,Kartaltepe2023,HuertasCompany2024,HuertasCompany2025}, and we discuss this further in Sec. \ref{sec:discussion}.

The star formation history of our low-mass progenitors is also mostly constant with redshift: most galaxies are star-forming regardless of their morphological type. There is a slight uptick in a quiescent population at the most recent times, with up to 20\% among the intermediate and irregular groups, consistent with late-time environmental quenching. There is also a slight increase of starbursts among the irregular galaxies in the highest redshift bin -- this could indicate a starburst driven by mergers or disk instabilities \citep[e.g.,][]{Ellison2008,Zolotov2015}, although the significance of this result is not sufficient to make any conclusions. Overall, we do not see a strong evidence for enhanced star formation among the irregular galaxies.

\subsubsection{Progenitors of massive galaxies}

\begin{figure*}
    \centering
    \includegraphics[width=\linewidth]{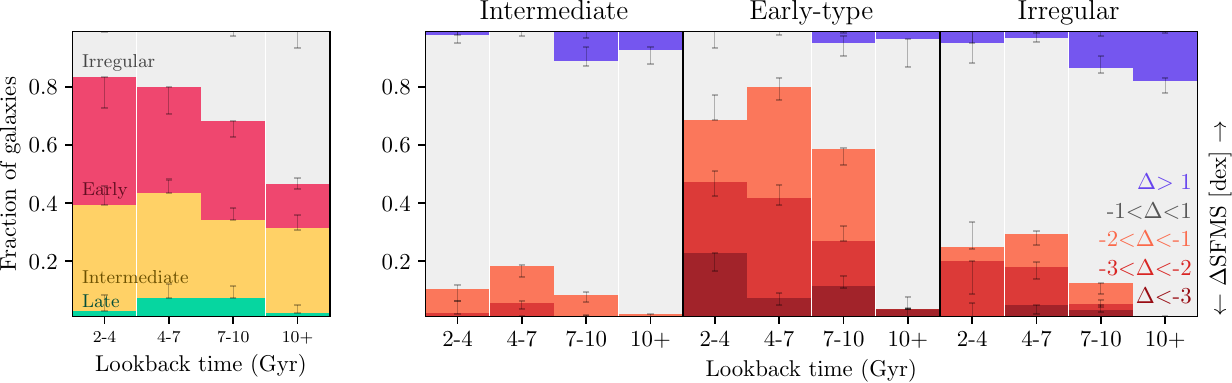}
    \caption{Same as Fig. \ref{fig:qfrac_lowm} for progenitors of more massive galaxies, $\log M_{\star,0} \in (10,11)$. While the fraction of LTGs is constant with redshift, the fraction of irregular galaxies decreases over time at the same time as the ETG fraction increases. LTGs are star-forming at all times, while ETGs star star-forming but quench rapidly. }
    \label{fig:qfrac_highm}
\end{figure*}

Fig. \ref{fig:qfrac_highm} shows the same plot for the progenitors of more massive galaxies, with $10 < \log M_{\star,0} < 11$. This is an interesting mass range since in the local Universe, galaxies in this mass range may be massive star-forming disks or quiescent ellipticals. We see a similar pattern in our most recent time bin: there is an even split of LTGs and ETGs (40\% each), with a small (20\%) fraction of irregular galaxies. 

There is a strong evolutionary trend in the massive galaxy progenitors. We see a constant fraction of LTGs, an increase in the number of irregulars, and a corresponding decrease in the number of ETGs with cosmic time. In the highest redshift bin, the fraction of irregular galaxies reaches $\sim$50\%, consistent with \cite{Kartaltepe2023} and \cite{HuertasCompany2024}. However, very quickly, in the next few Gyr this fraction drops to 25\% while the ETG fraction increases from $\sim$10\% to 30\%. 

A similar picture emerges when looking at star formation rates. The LTG population is predominantly star-forming at all cosmic times, with a slight uptick in quenching galaxies at later times. On the other hand, the ETGs start off almost entirely star-forming but quench very quickly, leading up to a 75\% fraction of galaxies off the star-forming main sequence at $z$$\sim$0.15. Our results support a picture where some galaxies in this mass range undergo a compaction event, going through a ``blue nugget'' and then a ``red nugget'' phase, becoming present-day ETGs; while other galaxies do not experience this and remain star forming disks at all times. We discuss this further in Sec. \ref{sec:discussion}.





\section{Discussion}\label{sec:discussion}

We now move on to our main question -- what is the evolution of galaxies across cosmic time in terms of mass, shape, and star formation? What is the relationship between these three quantities, and which bimodality -- mass-morphology, star formation-morphology, or mass-star formation, -- is the primary driver of galaxy evolution as a whole?

\subsection{The ``Hubble sequence'' and biases in morphological measurements}

When a ``Hubble sequence'' of galaxy morphologies first established is one of the cornerstone questions of galaxy evolution, and it is still disputed today. Earlier \textit{HST}-based studies suggested that a clean split between LTGs and ETGs disappears by $z\sim2$ \citep[e.g.,][]{Mortlock2013}, but JWST data has revealed that this may not be the case. Several studies have found that there is a significant population of apparently stable disks already present at a high redshift \citep{Ferreira2022a,Ferreira2022b,Sun2024,Shuntov2025,Costantin2025}. Studies based on visual or CNN-based classification also find a large number of ETG or spheroidal galaxies, especially at the high-mass end \citep{Kartaltepe2023,HuertasCompany2024,HuertasCompany2025}. Therefore, the consensus is increasingly becoming that the ``Hubble sequence'' and a mass-morphology relationship are already established by $z\sim3$. Our results support this picture: there is a clear ``main sequence'' in the UMAP space spanning low-mass, late-type galaxies to massive, early-type ones. We do not see a redshift gradient on this main sequence, which means that the main sequence galaxies are morphologically similar regardless of their redshift, and so the ``Hubble sequence'' does not change with lookback time. In other words, LTGs and ETGs are morphologically the same at $z$$\sim$0.15 and $z$$\sim$4.5, at least in the parameter space we studied. Therefore, we also agree that a ``Hubble sequence'' exists up to lookback times of 12.5 Gyr, or $z\sim4$. We also see an increasing number of ``irregular'' galaxies in our highest-redshift bin, which agrees with \cite{HuertasCompany2025} that it is possible that the ``Hubble sequence'' is first established in the 3$<$$z$$<$5 range. 

Quantitative studies of galaxy structure, however, paint a contradictory picture to the visual classifications. Visually classified LTGs and ETGs appear to have similar S\'ersic indices \citep{Ferreira2022b,Kartaltepe2023,Sun2024,Shuntov2025} and $G$--$M_{20}$ measurements \citep{Ferreira2022b,Costantin2025} that are most consistent with disks. The conclusion of these works has been that high-redshift early-type galaxies are not as centrally concentrated as their local counterparts.

\new{This is a contradiction. For visual classifications of local galaxies, scientists can look for substructures, such as spiral arms or star-forming knots, to distinguish LTGs and ETGs. However at high redshifts, where the effective resolution is insufficient to see spiral arms in most cases, classifiers primarily focus on how centrally concentrated the light is and the axis ratio of the object.We should therefore expect these classifications to agree with measurements of light concentration, and the apparent discrepancy between visual classes and S\'ersic indices is surprising.}

\new{This could be explained in two ways.} One reason for this apparent discrepancy could be the biases in structural measurements for sources at different distances. Several studies have shown that quantitative morphological parameters depend on image resolution and signal-to-noise ratio \citep[e.g.,][]{Graham2005,Holwerda2011,Bottrell2019,Ren2024,Sazonova2025}. Many works remedy this by redshifting simulated sources, applying cosmological surface brightness dimming and changing the angular size of the source, and then deriving average bias corrections for their metrics \citep[e.g.,][]{Whitney2021,Sun2024}. This is a good first approximation; but it does not account for an intrinsic size and luminosity evolution with redshift, which changes the \textit{effective} resolution of high-redshift sources. Effective resolution is an important factor in the performance of morphological metrics, and so it is better to account for this evolution in the redshifting procedure as was done in \cite{Ren2024}. However, one issue still remains: for many parameters the magnitude of the bias depends on the parameter's intrinsic value \citep{Thorp2021,Sazonova2025}, so applying an average correction does not fully account for the bias. \new{Another explanation may be that visual classifications are unreliable. If there is confusion between the two classes, their median S\'ersic indices will be similar.}

\new{Our work allows us to distinguish these two cases, since we carefully mapped all images to the same ``absolute'' resolution and surface brightness limit}. This way we are certain that any differences in structural parameters arise from \textit{intrinsic differences}, not measurement biases. After doing this exercise, we recovered a ``Hubble sequence'' even at high-redshift, and show that high-redshift LTGs and ETGs are similar to their low-redshift counterparts, at least up to our ability to see these differences. \new{We also showed that there is a significant population of centrally concentrated ETGs up to $z\sim4$.}

\new{We then compared S\'ersic index distributions of `ETG' and `LTG' classes from this work, from \cite{HuertasCompany2024},  and from \cite{Ferreira2023} (Fig. \ref{fig:morph_types}). While our ETGs have significantly higher S\'ersic indices than LTGs as expected, the visually classified ``disk'' and ``spheroid'' classes of \cite{Ferreira2023} have a significant overlap in their distribution of S\'ersic indices. Since we accounted for biases in measuring morphology, this favours the conclusion that the visual classifications are less reliable in distinguishing LTGs from ETGs than computational methods.}

\new{Interestingly, the deep learning classifications of \cite{HuertasCompany2024} have the expected S\'ersic index distributions, although roughly half of the ``spheroid'' galaxies have $n<2$ and so are likely misclassified LTGs.  These classifications are derived with a neural network trained on visual classifications of HST images with a domain adaptation step. In a way, this approach is similar to both visual and numeric classifications. The neural network computes image properties in its latent space -- akin to morphological parameters -- to output classifications, making it similar to our UMAP approach. However, the training set of the network is based on visual classifications, and so it is subject to all the biases present in the training set. One reason why it appears to perform better in distinguishing ETGs and LTGs could be the domain adaptation, which may make the model more robust to changes in image quality.}

\subsection{Low-mass progenitors: eternal disks}\label{sec:lowmass}

Fig. \ref{fig:qfrac_lowm}a) shows the relative fraction of ``early'', ``intermediate'', ``late'', and ``irregular'' galaxies (Fig. \ref{fig:morph_types}) as a function of cosmic time, for the progenitors of $9 < \log M_{\star,0} < 10$ (second row in Fig. \ref{fig:evolution}). Fig. \ref{fig:qfrac_lowm}b) shows the fraction of galaxies on the star-forming main sequence (grey), starbursts (purple), and quenching (shades of red) galaxies in the corresponding morphology bins. We do not show the ``early-type'' bin since the number of these are too low.

The progenitors of $\log M_{\star,0} \in (9, 10)$ galaxies are dominated by LTGs at all cosmic times (Fig. \ref{fig:qfrac_lowm}). In the local Universe, galaxies in this mass range are, for the most part, disks. \new{We see that their progenitors have the same quantitative morphology, which does not appear to evolve over time. This suggests that they are also disks at high redshift, agreeing} with previous studies, who found that galactic disks are established early and are already present by $z$$\sim$3 \citep{Ferreira2022a,Ferreira2022b,Kartaltepe2023,Lee2024}. \new{However, we can only conclude that LTGs do not evolve in terms of the features \textit{we can see}. In our matched resolution, most LTGs appear as featureless disks, even though we know that local galaxies have spiral arms, bars, and knots of star formation. It is possible that low-redshift LTGs have different substructures than their progenitors; but we would not be able to see this since we cannot resolve these structures at high $z$ even if they are there.}

Our LTGs are almost entirely star-forming at all cosmic times, as expected for low-mass galaxies \citep[e.g.,][]{Peng2010}. We see a slight increase in quenched fraction at lower lookback times, which appears independent of morphological type -- this is consistent with environmental quenching, where star formation shuts down due to gas removal from the galaxies' environment rather than a morphological evolution. Similar to \cite{HuertasCompany2025}, who found the same trend, we do not have environmental information to confirm this -- but our results agree with studies that find that low-mass galaxies are preferentially quenched in clusters since $z$$\sim$1 \citep[e.g.,][]{Socolovsky2018,Gully2025}. 

We find that the irregular fraction is constant with redshift ($\sim 25\%$), with a non-significant increase at high lookback times. This could imply that lower-mass progenitors have a quieter formation history, where they form a disk and experience relatively few mergers, allowing them to retain their disk morphology. Galaxies with a more violent merger history would already have a higher mass, and so would not be in this progenitor mass bin. However, there is one big caveat in our analysis: our high- and low-$z$ sources appear the same \textit{up to our image quality}. Since we matched the surface brightness limit of all images, the signal-to-noise ratio of our low-mass sample is low. We expect merger features of two low-mass galaxies to be fainter than of two massive ones, and they are almost definitely too shallow for us to detect in our imaging. Hence, we can conclude that there are few bright, trainwreck-type mergers among the low-mass progenitors, but there may be minor mergers we do not detect. 

There are several recent studies that report an intrinsic difference between low-z and high-z LTGs. \cite{Smethurst2025} and \cite{EspejoSalcedo2025}, using visual classification, show that from 50\% up to 75\% of high-redshift disks may be featureless, missing substructures such as spiral arms. Other works find an increasing number of clumpy disks compared to the local Universe, peaking at z$\sim$2 \citep[e.g.,][]{Sok2025}. Finally, \cite{Pandya2024} and \cite{VegaFerrero2024} suggested that there is a significant fraction of high-redshift sources that are intrinsically prolate and are just misclassified as edge-on disks. As we stated, we do not see a morphological difference between low-z and high-z LTGs in our UMAP phase space. In part, this is a strong limitation of our analysis: similar to our inability to detect merger features, due to our resolution limit, we would not detect clumps other than the brightest ones (these galaxies may be the ``irregular'' disks we found). However, this also applies to other observational studies. When we degraded the low-z imaging to match the high-z galaxies, local disks appear similarly featureless. While the JWST resolution is sufficient to study the global structure of low-mass galaxies, features such as spiral arms and bars have a much smaller typical size along one of their dimensions, and so are smoothed over by the PSF. Detecting these small features if local galaxies were moved to a higher redshift would be extremely challenging, as seen in Fig. \ref{fig:augments_comp} -- but perhaps possible by looking at S\'ersic residuals. 

\subsection{High-mass progenitors: two paths}\label{sec:highmass}

Fig. \ref{fig:qfrac_highm} shows a similar breakdown of galaxies by morphological type and star formation properties to Fig. \ref{fig:qfrac_lowm}, for $\log M_{\star,0} \in (10, 11)$ progenitor galaxies.

These galaxies tell a more complex story. We see three distinct groups in Fig. \ref{fig:qfrac_highm}: late-type, early-type, and irregular galaxies. Their relative fractions are strongly evolving: irregular galaxies are dominant ($\sim$50\%) at the earliest epoch, and over time become less common as the ETG population grows. Similar to \cite{Kartaltepe2023,HuertasCompany2024,HuertasCompany2025}, we find a significant ($\sim$30\%) fraction of ETGs already present by $z$$\sim$2. We also note that our sample of ETGs is likely not complete, since we excluded unresolved galaxies from our analysis -- so the most compact ETGs, such as little red dots \citep{Matthee2024} or post-starbursts \citep{Almaini2017}. The fraction of LTGs stays constant over our redshift range, again consistent with other studies. The abundance of both LTGs and ETGs is expected in this mass bin: $\log M_{\star,0} \in (10, 11)$ is an interesting mass range, where a galaxy with this stellar mass may be a massive spiral galaxy akin to the Milky Way, or may be a quiescent ETG. Our results thus agree that the progenitors of these galaxies can follow either one of those two paths. 

The relative fractions of the progenitor population let us speculate on the formation pathway of LTGs and ETGs in this mass range. The fraction of LTGs is constant over time, while the fraction of ETGs is increasing as the fraction of irregulars is decreasing -- this allows three possible interpretations.

A classic evolutionary picture is that massive galaxies first form disks, then undergo a compaction event that forms a spheroid, and then evolve passively via minor mergers into ETGs  \citep[e.g.,][]{Hopkins2006,Bournaud2007,Hopkins2008}. If this was the case, we would see a decrease in the LTG fraction as the ETG fraction increases, but we do not see this. It is possible that the irregular galaxies first become LTGs, which then become ETGs, \new{but the timescales for this change would have to be rapid, and so our data do not favour this model.}


\new{Our interpretation is that} similar to their lower-mass counterparts, \new{many} massive disks do not significantly evolve over time. They passively accrete gas and form stars, increasing in stellar mass, while retaining their disk morphology. This agrees with \cite{Tan2024}, who used abundance matching and found that Milky Way progenitors are consistent with star-forming disks throughout cosmic time. The early-type population is then \textit{not} built up from merging disks (at least at the z$<$4 redshift range probed here) -- instead, they start as galaxies with irregular morphologies, which quickly build up a central spheroid, and become early-type galaxies. ETGs are then an established population at $z$$<$3. \new{This is similar to the ``classic'' picture, except the early progenitors have irregular rather than disk structure at $z\sim4$.}

When they first form, ETGs are predominantly star-forming, but quench quickly, on timescales of a few Gyr. This picture agrees completely with the findings of \cite{HuertasCompany2024,HuertasCompany2025}, and a ``blue nugget'' -- ``red nugget'' formation scenario \citep{Barro2013,Dekel2013,HuertasCompany2018}, where massive galaxies first undergo compaction and then quench \citep[e.g.,][]{Tacchella2016,Lapiner2023,Tarrasse2025}. Based on the decreasing number of irregular galaxies as the ETG population becomes more dominant, the trigger of this compaction could be mergers, as predicted by hierarchical structure formation theory \citep[e.g.,][]{Springel2005,Hopkins2006,RodriguezGomez2015}. However, the compaction could also be caused by a disk instability, as suggested by \cite{Dekel2013} and \cite{Zolotov2015}. These disk instabilities can trigger large star-forming clumps, as is indeed observed in high-redshift galaxies \citep[e.g.,][]{Mowla2024,Sok2025}. Sufficiently bright clumps would also lead to high disturbance measurements and would be difficult to distinguish from mergers without color information. One way to determine whether these irregular galaxies undergo disk compaction or mergers is to look for close pairs \citep[e.g.,][]{Mantha2018,Chamberlain2024} -- if evolution is driven primarily by mergers, then we should expect to find a comparable number of close pair galaxies of this mass range just before they merge.

A third explanation could be that the ETGs we observe are not truly ETGs. While they are centrally concentrated and have high S\'ersic indices, many of these massive galaxies have a visible disk (Fig. \ref{fig:main_sequence}). One more explanation for this centrally concentrated light could be a presence of an AGN. We specifically excluded unresolved sources from our analysis, since they are likely to be AGN and quasars, which means we cannot reliably measure their morphology (or match our resolution requirements). The discovery of little red dots \citep[LRDs;][]{Matthee2024} revealed that at least some high-redshift galaxies have \textit{over-}massive black holes, with extremely bright and centrally concentrated optical emission coupled with a lack of X-ray, IR, or radio counterparts. While unresolved in the optical, these LRDs have extended UV profiles \citep{Chen2025,Zhang2025,Rinaldi2025}. Recently, \cite{Sun2026} showed that LRDs are well-modelled by a bright ``black hole star'' source \citep{Naidu2025a}, which emits in rest-frame optical and is superimposed onto a host galaxy. In particular, \cite{Sun2026} suggested that these objects may be much more common -- if a ``black hole star'' contribution is weaker compared to the underlying host, a galaxy may not be flagged as an LRD, and so a true abundance of these objects is still unknown. Our analysis focused on rest-frame optical morphology, exactly where the optical emission from such a central engine could dominate the light of the host galaxy. It is therefore possible that the high central concentrations we see are caused by an AGN (or a more exotic ``black hole star'') and not by a centrally concentrated stellar profile. 

If this were to be the case, we would expect that when the AGN duty cycle switches off, our ``ETGs'' would move back to the LTG category, replenishing the LTG population. This could allow a two-way evolutionary pathway where LTGs evolve into ETGs via mergers or compaction, and high-z ``ETGs'' (or AGN-dominated sources) evolve into LTGs as the AGN switches off. However, our star formation timescales make this scenario unlikely. We see an extremely rapid build-up of quiescent ETGs in our second latest time -- this is more consistent with rapid quenching post compaction than a longer ETG $\rightarrow$ LTG $\rightarrow$ Irr $\rightarrow$ ETG evolution. Moreover, we would expect such a strong AGN to quench star formation at least temporarily, but we do not see a significant quiescent LTG population in our sample. Therefore, we believe that while some of our ``ETG'' may be hidden AGN, they should comprise a relatively small fraction of our sample. It would be possible to clearly distinguish the two scenarios with a UV or IR follow-up.


\section{Summary}\label{sec:summary}

In this work, we created ``absolute'' images of galaxies spanning $0.15 < z < 4.5$ and $8 < \log M_\star < 11$, where the effective resolution and surface brightness limit were matched across the entire sample to allow consistent morphology measurements. We measured over 50 parameters with \texttt{statmorph-lsst} and \textsc{Galfit}, and used 24 of them to construct a morphological phase space using UMAP dimensionality reduction. 

Looking at this phase space, we learnt that:

\begin{enumerate}
    \item The mass-morphology ``main sequence'' is not bimodal, representing a continuum of galaxy structures, and exists at all cosmic times up to $z\sim4.5$ (Sec. \ref{sec:umap});
    \item At matched resolution and depth, most galaxies -- especially at the low mass -- are featureless regardless of redshift, since the resolution is not sufficient to detect faint substructures such as spiral arms (Sec. \ref{sec:umap});
    \item The UMAP-based morphological classifications perform similarly to deep learning and outperform simple visual classifications in terms of confusing LTGs and ETGs (Sec. \ref{sec:classifications});
    \item There is a significant and a heterogeneous ``irregular'' population with disturbed morphologies, potentially mapping different merger stages and mass ratios (Sec. \ref{sec:irregulars}).
\end{enumerate}

We then used empirical mass accretion history estimates to link the galaxies in our sample at lower redshifts to their likely progenitors, and looked at the evolution of galaxies' morphology over cosmic time. We found that:

\begin{enumerate}
\item Progenitors of low-mass galaxies have the same morphology and star formation properties at all cosmic times, showing a lack of evolution. While there may be some evolution in substructures (e.g. spiral arms), the resolution and depth of JWST is not sufficient to see it consistently at all redshifts (Sec. \ref{sec:lowmass}).

\item Progenitors of massive galaxies follow two paths: LTGs remain star-forming LTGs at all cosmic times, while ETGs show significant evolution (Sec. \ref{sec:highmass}).

\item The fraction of massive ETGs decreases with redshift at the same rate as the irregular fraction increases. The progenitors of massive ETGs are therefore likely irregular galaxies that undergo a compaction, form an ETG. ETGs start off star-forming but quickly quench star formation, favouring ``blue nugget'' -- ``red nugget'' scenario (Sec. \ref{sec:highmass}).

\end{enumerate}

The nature of the irregular progenitors of ETGs still remains uncertain in our analysis. It may be that beyond $z>4$ all galaxies have an intrinsically disturbed structure, they may be mergers, or extremely clumpy disk galaxies. 

Our analysis of the ``irregular'' population on the UMAP reveals that there are distinct classes of ``irregular'' galaxies that may have a different origin. A comparison to simulated galaxies with known merger stages and mass ratios may help to map UMAP regions to a merger sequence. Similarly, parts of the region may correspond to clumpy or galaxies with irregular shapes. We will explore these tests in a future follow-up. 

Another obvious pathway to study the nature of the irregular ETG progenitors is to look at higher redshifts. However, we used a specific redshift cut-off as cosmological surface brightness dimming makes it challenging to find disturbed features beyond $z>4$ in the existing JWST surveys. Deeper surveys, such as JWST JADES \citep{jades} or an ultra-deep survey akin to the Hubble eXtreme deep field, may help remedy this problem. With existing data, machine learning techniques may help us recover faint features that are lost to cosmological dimming.

\section*{Acknowledgements}

We thank the University of Waterloo for their support of the Canada Rubin Fellowship program. ES thanks Roan Haggar, Ana Ennis, and the members of the unofficial ``galaxies office'' for many fruitful scientific discussions throughout this project. CRM acknowledges support from an Ontario Graduate Scholarship. MLB acknowledges support from an NSERC Discovery Grant.

\vspace{5mm}
\textit{Facilities:} JWST (NIRCam), HST (WFC3, ACS), MAST, HLA, NED

\vspace{5mm}
\textit{Software:} \texttt{statmorph-lsst} \citep{statmorph,Sazonova2025}, photutils \citep{photutils}, astropy \citep{astropy,astropy2,astropy3}, reproject \citep{reproject}, \textsc{Galfit} \citep{galfit,galfit2}, Scikit-Image \citep{skimage}, Scikit-Learn \citep{sklearn}, Matplotlib \citep{matplotlib}, pandas \citep{pandas}, NumPy \citep{numpy}, SciPy \citep{scipy}

\bibliography{references}{}
\bibliographystyle{mnras_custom}

\clearpage
\appendix

\section{Structural parameters}\label{app:params}

Table \ref{tab:parameters} lists the primary structural parameters computed by \texttt{statmorph-lsst}, their descriptions, and relevant references. We used a subset of 24 parameters in our UMAP analysis, indicated by the ``Used'' colummn. For the parameters we omitted, we briefly state the reason. 

{\renewcommand{\arraystretch}{1.5}
\begin{longtable*}{p{0.07\textwidth} p{0.12\textwidth} p{0.58\textwidth} p{0.12\textwidth} p{0.03\textwidth}}

\caption{Structural parameters measured from galaxy images}\label{tab:structural_params}\\
\toprule
Symbol & Param. name & Description & Refs. & Used\\
\midrule
\midrule

\multicolumn{5}{l}{\textit{Geometric measurements}}\\
\midrule
$(x_0, y_0)$ & \texttt{xc\_centroid,} \texttt{yc\_centroid} &
    Centroid of the galaxy light distribution computed from image moments. \newline Not used since it doesn't carry morphological information.  &
    \citetalias{statmorph,Sazonova2025} & \\
    
$e$ & \texttt{ellipticity} \texttt{\_centroid} &
    Ellipticity measured from the image moments about the centroid ($e = 1 - b/a$). &
    \citetalias{statmorph,Sazonova2025} & \checkmark\\

$\varepsilon_c$ & \texttt{elongation\_} \texttt{centroid} &
    Elongation ($a/b$) derived from ellipticity. Not used since redundant with ellipticity. &
    \citetalias{statmorph} & \\

$\theta_c$ & \texttt{orientation\_} \texttt{centroid} &
    Position angle of the major axis measured about the centroid, measured from the image moments. Not used since orientations are random. &
    \citetalias{statmorph} & \\

$(x_A, y_A)$ & \texttt{xc\_asymmetry,} \texttt{yc\_asymmetry} &
    Centre of rotation that minimizes the RMS asymmetry $A_{\rm RMS}$. Not used.  &
    \citetalias{statmorph, Conselice2003,Sazonova2025} & \\

$e_A$ & \texttt{ellipticity\_} \texttt{asymmetry} &
    Same as $e$, measured from image moments about $(x_A, y_A)$. \newline 
    Not used since we already use $e$ in the UMAP. &
    \citetalias{statmorph} & \\

$\varepsilon_A$ & \texttt{elongation\_} \texttt{asymmetry} &
    Same as $\varepsilon$, measured about $(x_A, y_A)$.  &
    \citetalias{statmorph} & \\

$\theta_A$ & \texttt{orientation\_} \texttt{asymmetry} &
    Same as $\theta$, measured about $(x_A, y_A)$. &
    \citetalias{statmorph} & \\

\midrule
\multicolumn{5}{l}{\textit{Flux measurements} -- do not capture morphology and not used.}\\
\midrule
$F_{\rm circ}$ & \texttt{flux\_circ} &
    Total flux of the galaxy within 1.5 times the circular Petrosian radius $R_{p,\circ}$ &
    \citetalias{statmorph}, \citetalias{Petrosian1976} & \\

$F_{\rm ellip}$ & \texttt{flux\_ellip} &
    Total flux of the galaxy within 1.5 times the elliptical Petrosian radius $R_{p,e}$ &
    \citetalias{statmorph}, \citetalias{Petrosian1976} & \\

\midrule
\multicolumn{5}{l}{\textit{Radius measurements} -- not used since we did not want mass-size relation to dominate the morphological distribution.}\\
\midrule
$R_{p,\circ}$ & \texttt{rpetro\_circ} &
    Circular Petrosian radius: the radius at which the mean surface brightness in an annulus equals $\eta = 0.2$ times the mean surface brightness within that radius. &
    \citetalias{Petrosian1976,statmorph,Sazonova2025} & \\

$R_{p,e}$ & \texttt{rpetro\_ellip} &
    Elliptical Petrosian radius, computed using elliptical isophotes whose shape is fixed to the ellipticity and orientation from image moments. &
    \citetalias{Petrosian1976,statmorph,Sazonova2025} & \\

$R_{\rm max,\circ}$ & \texttt{rmax\_circ} &
    Maximum radius of the segmentation footprint in a circular aperture. &
    \citetalias{statmorph} & \\

$R_{\rm max,e}$ & \texttt{rmax\_ellip} &
    Maximum semi-major axis of the segmentation footprint in an elliptical aperture. &
    \citetalias{statmorph} & \\

$R_{50,\circ}$ & \texttt{rhalf\_circ} &
    Circular half-light radius: radius enclosing 50\% of the total flux, where the total flux is the flux within $R_{\rm max,\circ}$. &
    \citetalias{Bershady2000,statmorph} & \\

$R_{50,e}$ & \texttt{rhalf\_ellip} &
    Elliptical half-light radius: semi-major axis of the elliptical aperture enclosing 50\% of the total flux, where the total flux is defined within an elliptical aperture with size $R_{\rm max,e}$, elongation $\epsilon_A$ and position angle $\theta_A$. &
    \citetalias{statmorph} & \\

$R_{20}$ & \texttt{r20} &
    Isophotal radius enclosing 20\% of the total galaxy flux, where the total flux is defined within a circular aperture with size $1.5R_p$.  &
    \citetalias{Bershady2000,Conselice2003} & \\

$R_{50}$ & \texttt{r50} &
    Same as $R_{20}$, for 50\% of the total flux. &
    \citetalias{Bershady2000} & \\

$R_{80}$ & \texttt{r80} &
    Same as $R_{20}$, for 80\% of the total flux. &
    \citetalias{Bershady2000,Conselice2003} & \\

\midrule
\multicolumn{5}{l}{\textit{Bulge strength measurements}}\\
\midrule
$G$ & \texttt{gini} &
    Gini coefficient of the galaxy pixel flux distribution; quantifies inequality in light distribution among pixels. Higher values indicate more concentrated light &
    \citetalias{Abraham2003}, \citetalias{Lotz2004} & \checkmark\\
 
$M_{20}$ & \texttt{m20} &
    Second-order moment of the brightest 20\% of the galaxy's flux, normalized by the total second-order moment. Traces spatial distribution of bright features &
    \citetalias{Lotz2004} & \checkmark\\
 
$B(G,M_{20})$ & \texttt{gini\_m20}      \texttt{\_bulge} &
    $G$--$M_{20}$ bulge statistic; a linear combination of $G$ and $M_{20}$ that separates bulge-dominated from disk-dominated galaxies &
    \citetalias{Lotz2004,statmorph} & \checkmark\\
 
$C$ & \texttt{concetration} &
    Concentration index, defined as $C = 5\log_{10}(R_{80}/R_{20})$, where $R_{80}$ and $R_{20}$ are the radii enclosing 80\% and 20\% of the total flux. &
    \citetalias{Kent1985,Bershady2000,Conselice2003} & \checkmark \\

$R_{20}/R_{50}$ & \texttt{r20\_50} &
    Ratio of the 20\%-light radius to the 50\%-light (half-light) radius; traces inner light profile concentration &
    \citetalias{Bershady2000} & \checkmark\\
 
$R_{80}/R_{50}$ & \texttt{r80\_50} &
    Ratio of the 80\%-light radius to the 50\%-light radius; traces the extent of the outer light profile relative to the core &
    \citetalias{Bershady2000} & \checkmark\\

\midrule
\multicolumn{5}{l}{\textit{Disturbance / merger indicators}}\\
\midrule
$S(G,M_{20})$ & \texttt{gini\_m20} \texttt{\_merger} &
    $G$--$M_{20}$ merger statistic; a linear combination of $G$ and $M_{20}$ designed to identify merging systems. &
    \citetalias{Lotz2004,statmorph} & \checkmark\\
 
$A_{\rm{CAS}}$ & \texttt{asymmetry} &
    Absolute (CAS) asymmetry, quantifying the degree to which a galaxy's light distribution is asymmetric under 180\degree\ rotation. &
    \citetalias{Conselice2000,Conselice2003,statmorph} & \checkmark\\
 
$A_{o}$ & \texttt{outer\_} \texttt{asymmetry} &
    Outer asymmetry, measured the same way as same $A_{\rm{CAS}}$ by excising the pixels within $R_{50,\circ}$. Sensitive to faint tidal features in galaxy outskirts. &
    \citetalias{Wen2016} & \checkmark\\
 
$A_{S}$ & \texttt{shape\_} \texttt{asymmetry} &
    Shape asymmetry, the asymmetry of the galaxy's binary detection footprint (segmentation map) rather than its flux distribution &
    \citetalias{Pawlik2016} & \checkmark\\

$A_{\rm RMS}$ & \texttt{arms} &
    RMS asymmetry: rotational asymmetry computed from the squared residual flux rather than the absolute value. &
    \citetalias{Conselice2003,Sazonova2024}. &  \checkmark \\

$A_{\rm{iso},\, X}$ & \texttt{aiso\_X} &
    Isophotal asymmetry: the asymmetry of a binary map detection map above a given surface brightness isophote $X$. For this work, this is an array with $X = 19, 19.5, 20, 20.5, 21, 21.5, 22, 22.5, 23$. Sensitive to disturbances of different surface brightness levels based on the value of $X$. 
    
    &
    \citetalias{Sazonova2025}& \checkmark\\

\midrule
\multicolumn{5}{l}{\textit{Other measurements}}\\
\midrule
$M$ & \texttt{multimode} &
    Multimode statistic; identifies galaxies with multiple modes (peaks) in their light distribution, sensitive to double nuclei and mergers. &
    \citetalias{Freeman2013} & \checkmark\\
 
$D$ & \texttt{deviation} &
    Deviation statistic; measures the distance between the brightest pixel and the galaxy centroid, normalized by the Petrosian radius. &
    \citetalias{Freeman2013} & \checkmark\\

$I$ & \texttt{intensity} &
    Intensity statistic; ratio of the total flux in the second-brightest to the brightest watershed segment of the smoothed image; sensitive to double-peaked light distributions. Not used as it is extremely sensitive to the watershed map and the results are unstable, especially in noisy imaging. &
    \citetalias{Freeman2013,Peth2016} & \\
 
$S$ & \texttt{smoothness} &
    Smoothness index; fraction of flux in high-spatial-frequency residuals after subtracting a smoothed image. Not used since it is dominated by noise in shallow imaging &
    \citetalias{Conselice2003} & \\

$St$ & \texttt{substructure} &
    Similar to $S$, with an additional step of running source detection on residual clumps to minimize noise contribution. &
    \citetalias{Sazonova2025} & \checkmark\\

\midrule
\multicolumn{5}{l}{\textit{S\'{e}rsic parameters} -- not used since not all S\'ersic fits were successful.}\\
\midrule
$n$ & \texttt{sersic\_n} &
    S\'{e}rsic index from a single-component PSF-convolved S\'{e}rsic fit; $n = 1$ corresponds to an exponential (disk) profile and $n = 4$ to a de Vaucouleurs elliptical profile. &
    \citetalias{Sersic1963,galfit} & \\

$R_{e}$ & \texttt{sersic\_rhalf} &
    Effective (half-light) radius of the best-fitting S\'{e}rsic profile. &
    \citetalias{Sersic1963,galfit} & \\

$I_0$ & \texttt{sersic\_} \texttt{amplitude} &
    Central intensity of the best-fitting S\'{e}rsic profile, in image flux units per pixel &
    \citetalias{Sersic1963,galfit} & \\

$(x_e, y_e)$ & \texttt{sersic\_xc,} \texttt{sersic\_yc} &
    Centre of the best-fitting S\'{e}rsic profile, in pixel coordinates within the postage stamp &
    \citetalias{galfit} & \\

$e_e$ & \texttt{sersic\_ellip} &
    Ellipticity ($1 - b/a$) of the best-fitting S\'{e}rsic profile; more robust to PSF effects than the non-parametric estimate &
    \citetalias{galfit} & \\

$\theta_e$ & \texttt{sersic\_theta} &
    Position angle of the best-fitting S\'{e}rsic profile, in radians &
    \citetalias{galfit} & \\

$\chi^2_\nu$ & \texttt{sersic\_chi2} \texttt{\_dof} &
    Reduced chi-squared of the S\'{e}rsic fit, measuring goodness of fit per degree of freedom &
    \citetalias{galfit} & \\

\label{tab:parameters}
\end{longtable*}
}

\section{Alternative ``absolute'' images}\label{app:simple_augments}

In the main body of the paper, we created ``absolute'' images by scaling the galaxies' flux to an ``absolute'' distance of 10 Mpc, matching effective resolution of all objects, and correcting for the evolution in the mass-to-light ratio. A simpler approach to the ``absolute'' images is, once again, move every source to a distance of 10 Mpc, resample all galaxies to the same resolution and pixel scale, and add noise to match the same surface brightness limit. This is the choice we explore here. 

We chose the resolution in pc/px, using the lowest baseline of our sample. For the pixel scale, this limit is set by the 0.03$\arcsec$ pixel scale at $z=1.5$ where the physical resolution is 260 pc/px. Beyond this redshift, the angular diameter scale turns over and the resolution begins to improve. Our limiting PSF FWHM is 1080 pc, set by the wider HST PSF ($z=0.815$, F814W filter). We convolved all images to match a 1080pc PSF FWHM and downsampled them to a 260 pc/px pixel scale. 

In this approach, we ignored the evolution in the mass-to-light ratio and the fact that the high-redshift sources are intrinsically brighter. After moving all sources to a flux distance of 10 Mpc, the 90\ts{th} percentile of surface brightness limits for this sample was 21.7 mag/arcsec$^2$. We then added noise to match this surface brightness limit as before. 

\begin{figure}
    \centering
    \includegraphics[width=\linewidth]{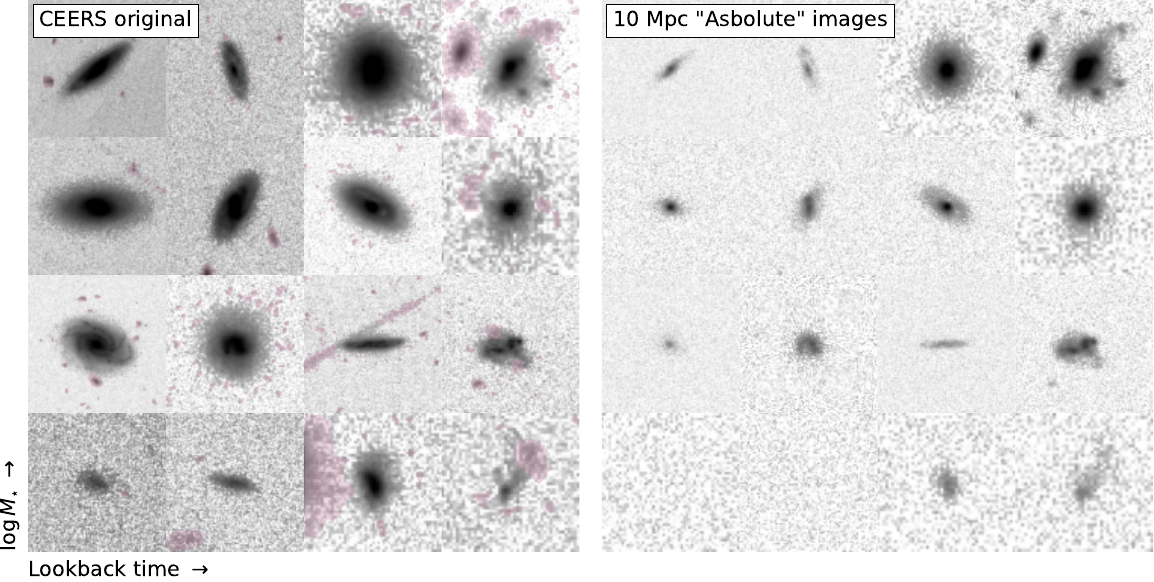}
    \caption{Same as Fig. \ref{fig:augments}, using simpler image quality matching, where all galaxies are resampled to the same 260 pc/px pixel scale with 1080pc PSF FWHM and 21.7 mag/arcsec$^2$ 10 Mpc surface brightness limit.}
    \label{fig:augments_simple}
\end{figure}

There are two crucial differences between this approach and that taken in the main body of the paper. First, since the resolution is set on a physical scale and high-redshift sources are intrinsically smaller, the \textit{effective} resolution of augmented images is lower for high-z galaxies. Second, since we did not account for the mass-to-light ratio evolution, the \avgsnr is lower for nearby galaxies at our shallower surface brightness limit. Fig. \ref{fig:augments_simple} shows a comparison of the original and ``absolute'' images with this approach -- the images of local and distant sources are clearly distinguishable by eye since the high-redshift galaxy images are smaller and brighter.

\begin{figure}
    \centering
    \includegraphics[width=\linewidth]{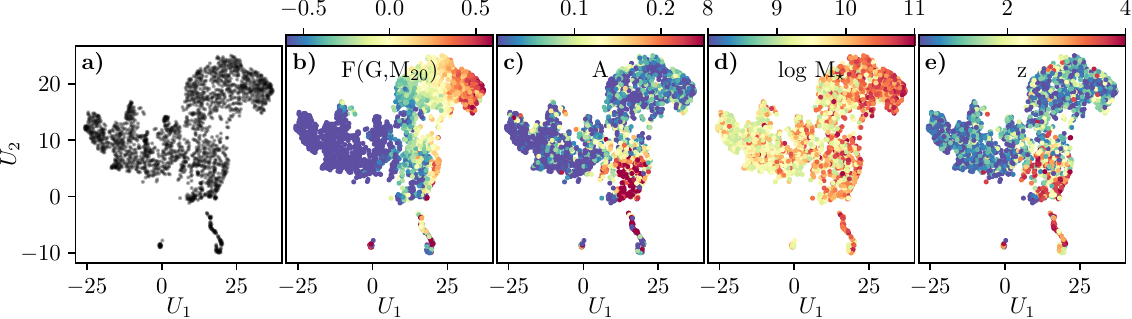}
    \caption{Same as Fig. \ref{fig:umap}, for a UMAP distribution derived using the alternative ``absolute'' images. Panels b) and c) show the distribution of the Gini-M$_{20}$ bulge strength and asymmetry, respectively. Panels c) and d) show stellar mass and redshift, which were not included in the UMAP.}
    \label{fig:umap_simple}
\end{figure}

We measured the morphologies of this set of images and passed the same parameters through a UMAP algorithm, same as in Sec. \ref{sec:umap}. Fig. \ref{fig:umap_simple}a) shows the UMAP distribution for the non-cosmological approach. In panels b, c, d, and e we colored the points by their bulge strength, asymmetry, stellar mass, and redshift respectively. 

The same mass-morphology ``Hubble sequence'' gradient is present in the new UMAP, tracing galaxies from late-type on the left to early-type on the right. The main difference is that now, the high-redshift sources occupy a distinct region of the UMAP off the ``Hubble sequence'' due to their high asymmetry.

It is tempting to interpret this distribution as evidence for the fact that high-redshift galaxies have a fundamentally different structure from their low-redshift descendants, and form an irregular population inconsistent with the local ``Hubble sequence''. However, asymmetry metrics in particular are extremely sensitive to the signal-to-noise ratio. High-redshift galaxies are intrinsically brighter, and given our extremely shallow surface brightness limit, their irregular features are more detectable. This apparent increase in asymmetry should be attributed to an inconsistent image quality between the low- and the high-redshift samples, and confirms that our main approach is better at mitigating the biases in morphological measurements. Since the morphological measurements derived with this approach are less reliable, we did not continue with the rest of the analysis.

\section{UMAP + S\'ersic parameters}\label{app:sersic_umap}

\begin{figure}
    \centering
    \includegraphics[width=\linewidth]{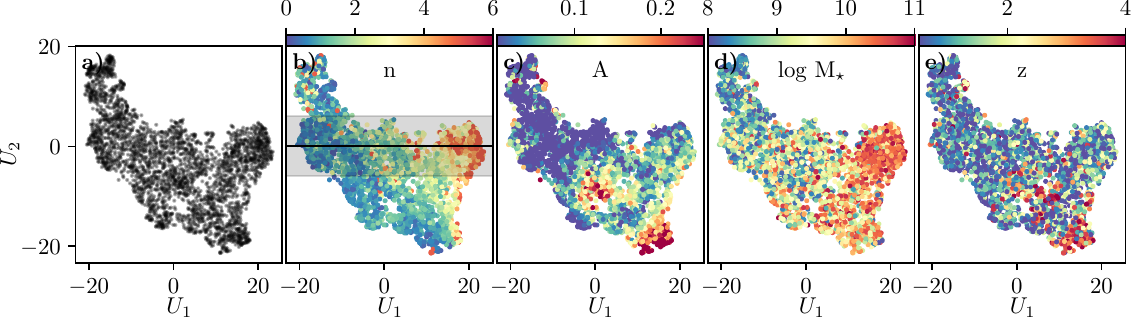}
    \caption{Same as Fig. \ref{fig:umap_simple}, for a UMAP distribution with cosmological augmentations and including the S\'ersic index. There is still a clear ``mass-morphology main sequence'', outlined as a grey line and a shaded region in panel \textbf{b)}.}
    \label{fig:umap_sersic}
\end{figure}

In Fig. \ref{fig:umap_sersic}, we show the UMAP produced after including the S\'ersic index and ellipticity into the UMAP fit. Qualitatively, the distribution is similar to Fig. \ref{fig:umap}, with a distinct mass-morphology sequence spanning from late-type low-mass galaxies to early-type massive ones. In this iteration of the UMAP, we can simply approximate this main sequence as a horizontal line with $U_1 \in (-6, 6)$ (grey shaded region).
\begin{wrapfigure}{r}{0.4\textwidth}
    \centering
    \includegraphics[width=0.4\textwidth]{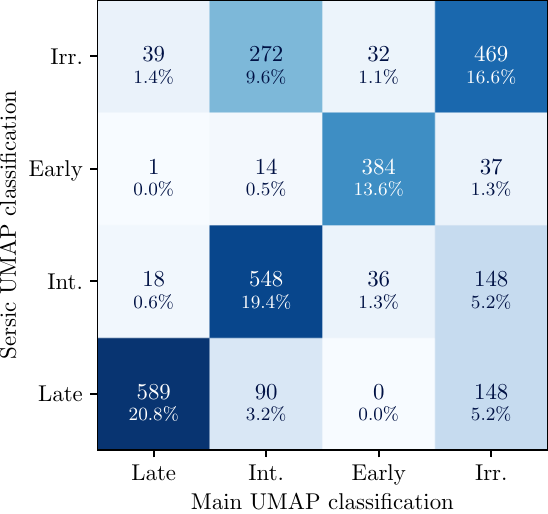}
    \caption{}
    \label{fig:confmat}
    \vspace{-5mm}
\end{wrapfigure}

The large cluster of low-$n$ galaxies consist of low \avgsnr{} images, where the galaxies are only marginally detected. Although they lie off the ``main sequence'', they form a distinct cluster in the irregular UMAP space, and so can be included in the ``late-type'' category (as we did in Sec. \ref{sec:irregulars}). Repeating the same steps with this new UMAP space as in the main body of the paper, we recover the same conclusions as in Figures \ref{fig:evolution}, \ref{fig:qfrac_lowm}, and \ref{fig:qfrac_highm}, and so we do not include these plots. It is reassuring, however, that our results with and without the S\'ersic fit agree.

In terms of morphological classifications based on the UMAP space, the S\'ersic approach agrees with 70\% of our classifications overall. Fig. \ref{fig:confmat} shows the confusion matrix of the four classes using the two approaches. The agreement is good overall, with the most discrepancy between the intermediate and irregular classes using the two approaches. Inspecting the misclassified galaxies visually, it is not apparent whether one approach is more pure or complete than the other: there are irregular galaxies that we classified as ``intermediate'' and vice versa. Overall, it is challenging to build a pure and complete sample of merging galaxies, and we note that the two methods give a roughly 10\% disagreement. Since S\'ersic fits failed for a significant fraction of our galaxies ($\sim$10\%), we opted to focus on the fully non-parametric UMAP in our analysis.

\section{Parameter distributions}\label{app:parameter_dist}

\begin{figure}[b]
    \centering
    \includegraphics[width=\linewidth]{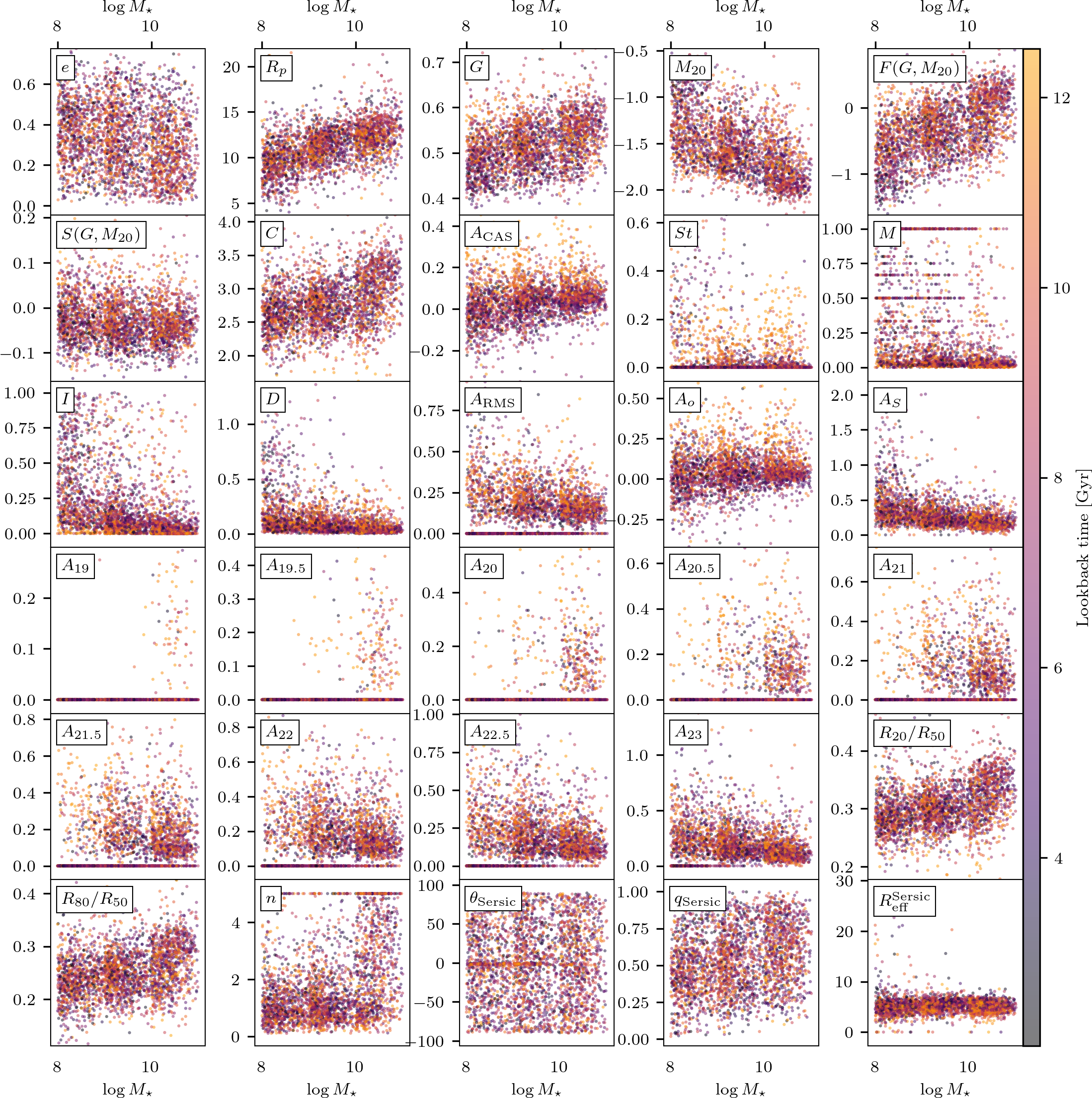}
    \caption{The distributions of 30 morphological parameters of \texttt{statmorph-lsst}, listed in Table \ref{tab:parameters}, as a function of stellar mass and lookback time.}
    \label{fig:parameters}
\end{figure}

\begin{figure}
    \centering
    \includegraphics[width=0.93\linewidth]{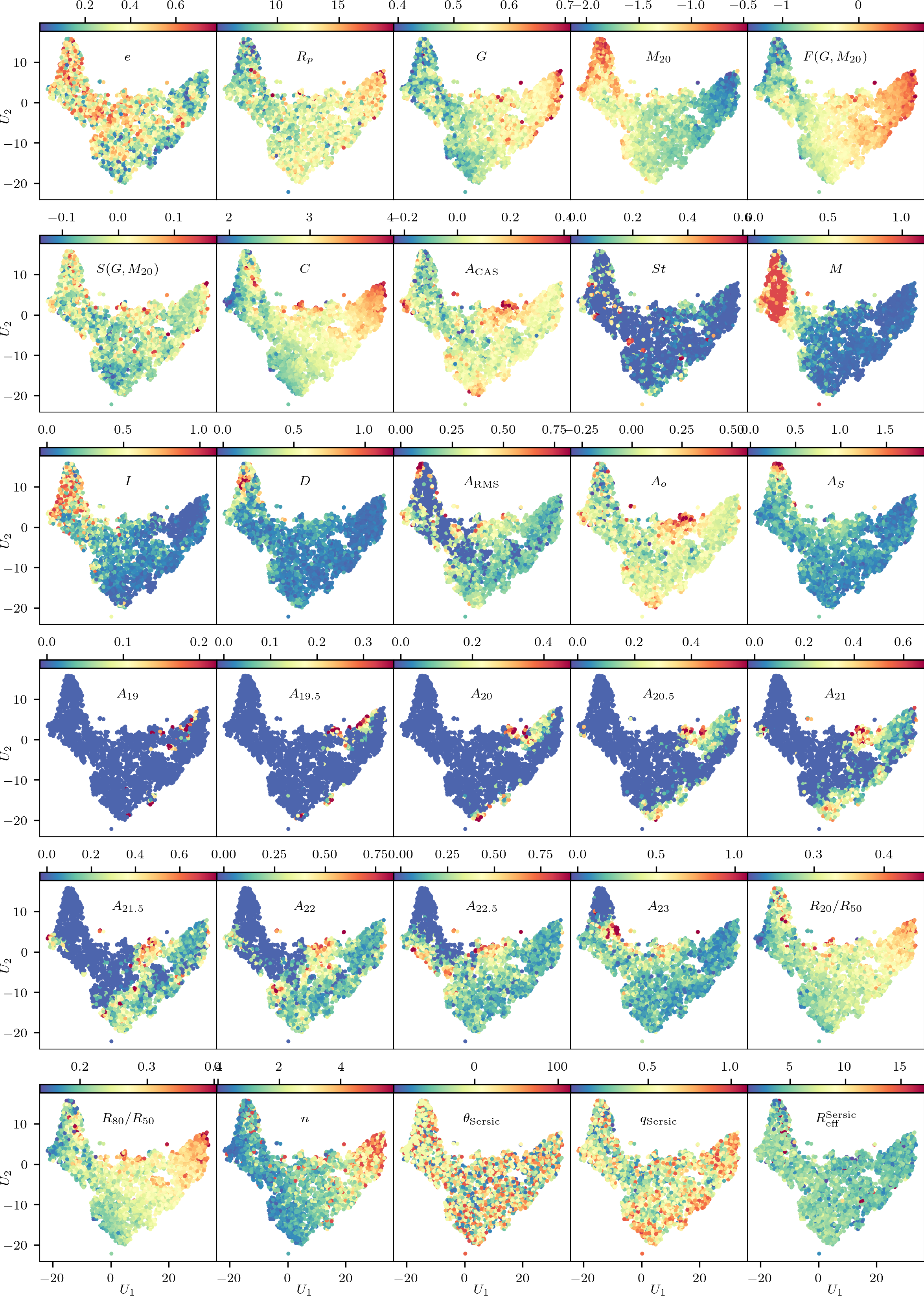}
    \caption{Same as Fig. \ref{fig:umap}, where each point in the UMAP phase space is colored by its parameter value from Fig. \ref{fig:parameters}.}
    \label{fig:umap_all}
\end{figure}

Fig. \ref{fig:parameters} shows the distributions of morphological parameters measured with \texttt{statmorph-lsst} as a function of stellar mass and lookback time for reference. Due to large dimensionality of this space, it is much more challenging to see trends in this data, justifying our use of the UMAP method. Fig. \ref{fig:umap_all} reproduces Fig. \ref{fig:umap}, coloring the phase-space points by a full range of parameters. Overall the distributions agree with the representative parameters we chose in Fig. \ref{fig:umap}.

\end{document}